\documentclass[manuscript]{aastex63}
\usepackage{graphicx}
\usepackage{supertabular}
\usepackage{longtable}
\usepackage{CJK}
\usepackage{tikz}
\usepackage{threeparttable}
\usetikzlibrary{shapes,arrows}
\received{----}
\revised{----}
\accepted{November 17, 2021}
\submitjournal{ApJS}

\begin{document}
\begin{CJK*}{UTF8}{gbsn}
\title{Dust Extinction Law in Nearby Star-Resolved Galaxies. I. M31 Traced by Supergiants}

\correspondingauthor{Jian Gao}
\email{jiangao@bnu.edu.cn\\yiren@mail.bnu.edu.cn}

\author[0000-0003-3860-5286]{Yuxi Wang (王钰溪)}
\affiliation{Department of Astronomy, Beijing Normal University, Beijing 100875, People's Republic of China; \rm{\href{jiangao\@bnu.edu.cn}{jiangao@bnu.edu.cn}}}

\author[0000-0003-4195-0195]{Jian Gao (高健)}
\affiliation{Department of Astronomy, Beijing Normal University, Beijing 100875, People's Republic of China; \rm{\href{jiangao\@bnu.edu.cn}{jiangao@bnu.edu.cn}}}

\author[0000-0003-1218-8699]{Yi Ren (任逸)}
\affiliation{Department of Astronomy, Beijing Normal University, Beijing 100875, People's Republic of China; \rm{\href{jiangao\@bnu.edu.cn}{jiangao@bnu.edu.cn}}}
\affiliation{College of Physics and Electronic Engineering, Qilu Normal University, Jinan 250200, People's Republic of China; \rm{\href{yiren\@mail.bnu.edu.cn}{yiren@mail.bnu.edu.cn}}}



\begin{abstract}
The dust extinction laws and dust properties in M31 are explored with a sample of reddened O-type and B-type supergiants obtained from the LGGS.
The observed spectral energy distributions (SEDs) for each tracer are constructed with multiband photometry from the LGGS, PS1 Survey, UKIRT, PHAT Survey, Swift/UVOT and XMM-SUSS.
We model the SED for each tracer in combination with the intrinsic spectrum obtained from the stellar model atmosphere extinguished by the model extinction curves.
Instead of mathematically parameterizing the extinction functions, the model extinction curves in this work are directly derived from the silicate-graphite dust model with a dust size distribution of $dn/da \sim a^{-\alpha}{\rm exp}(-a/0.25),~0.005 < a < 5~\mu {\rm m}$.
The extinction tracers are distributed along the arms in M31, with the derived MW-type extinction curves covering a wide range of $R_V$ ($\approx 2 - 6$), indicating the complexity of the interstellar environment and the inhomogeneous distribution of interstellar dust in M31.
The average extinction curve with $R_V \approx 3.51$ and dust size distribution $dn/da \sim a^{-3.35}{\rm exp}(-a/0.25)$ is similar to those of the MW but rises slightly less steeply in the far-UV bands, implying that the overall interstellar environment in M31 resembles the diffuse region in the MW.
The extinction in the $V$ band of M31 is up to 3 mag, with a median value of $ A_V \approx 1$ mag.
The multiband extinction values from the UV to IR bands are also predicted for M31, which will provide a general extinction correction for future works.
\end{abstract}



\section{Introduction} \label{sec:intro}

Dust extinction is crucial for recovering the intrinsic spectral energy distributions (SEDs) of celestial objects and inferring the characteristics of interstellar dust.
The wavelength dependence of interstellar extinction (commonly known as the extinction law or extinction curve) is defined as $A_{\lambda}$ at wavelength $\lambda$.
Since the absolute value of extinction $A_{\lambda}$ is difficult to directly calculate, the relative extinction $A_{\lambda}/A_V$, or the color excess ratio $E(\lambda-V)/E(B-V)=(A_{\lambda}-A_V)/(A_B-A_V)$, is frequently used to indicate dust extinction.

\citet[hereafter CCM]{1989ApJ...345..245C} found that the dust extinction law in the Milky Way (MW) from ultraviolet (UV) to near infrared (IR) could be characterized by one parameter $R_V$ [$R_V = A_V/E(B-V)$], which is called the total-to-selective extinction ratio.
Typically, the value of $R_V$ depends on the interstellar environment along the sightline, indicating the dust size.
The galactic diffuse regions have an average value of $R_V \approx 3.1$ \citep{2003ARA&A..41..241D}.
In dense molecular clouds, $R_V$ could be as large as $\approx 6$ \citep{1999PASP..111...63F, 1990ARA&A..28...37M}, while it could be $\approx 2$ in low-density regions \citep{1999PASP..111...63F,2017ApJ...848..106W}.

\begin{figure}
	\centering
	\includegraphics[scale=0.7]{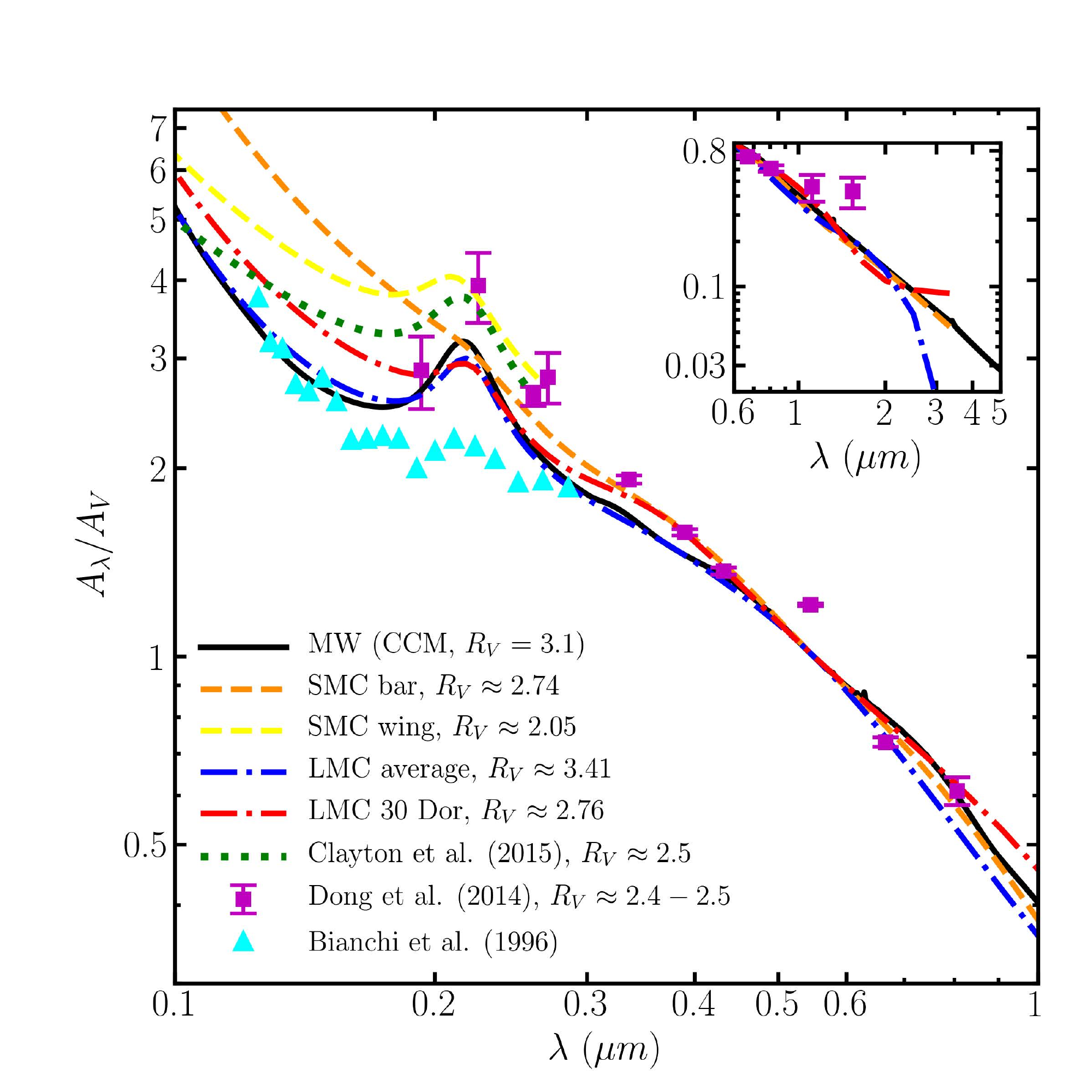}
	\caption{Extinction curves of the four local group galaxies: the MW, LMC, SMC and M31.
	It presents the extinction law from UV ($0.1~\mu {\rm m}$) to $1~\mu {\rm m}$, while the inset shows the extinction law from $0.6~\mu {\rm m}$ to IR bands ($5~\mu {\rm m}$).
	The black solid line shows the extinction law of the diffuse region in the MW.
    The orange dashed line and the yellow dashed line are the extinction curves in the SMC bar region and the SMC wing region, respectively \citep{2003ApJ...594..279G}.
	The blue dashed-and-dotted line presents the LMC average extinction law, while the red dashed-and-dotted line is the extinction curve of the 30 Dor region in the LMC \citep{2003ApJ...594..279G}.
	The cyan triangles show the MW-like extinction law with a weaker 2175 $\,{\rm \AA}$ bump derived by \citet{1996ApJ...471..203B}.
	The purple squares indicate the extinction near the bulge in M31 \citep{2014ApJ...785..136D}, while the green dotted line is the average of the four extinction curves derived by \citet{2015ApJ...815...14C}.
	\label{fig:M31_intro}
	The $R_V$ values of MCs are referred to \citet{2003ApJ...594..279G}.
	}
\end{figure}

However, an increasing number of studies on extinction beyond the MW have determined that the MW-type dust extinction law is not generally suitable for external galaxies.
As illustrated in Figure \ref{fig:M31_intro}, although the average Large Magellanic Cloud (LMC) extinction law \citep{1978Natur.276..478N,2003ApJ...594..279G} resembles that in the MW, the extinction law near the 30 Doradus star-forming region \citep{1985ApJ...288..558C,1986AJ.....92.1068F,2003ApJ...594..279G} has a very weak 2175 $\,{\rm \AA}$ bump.
In addition, most extinction curves in the Small Magellanic Cloud (SMC) bar region \citep{1984A&A...132..389P,2003ApJ...594..279G} display a nearly linear rise with wavelength $\lambda^{-1}$ and an absent 2175 $\,{\rm \AA}$ bump.
Meanwhile, the extinction curve towards AzV 456 located in the SMC wing region \citep{1982A&A...113L..15L,2003ApJ...594..279G} has a flatter slope and a weak 2175 $\,{\rm \AA}$ bump that cannot be interpreted by the single-parameter CCM function.

The Andromeda Galaxy, which is usually called M31, is the third nearest to us ($\approx 780~{\rm kpc}$, \citealt{2005MNRAS.356..979M}) after the LMC and SMC.
Individual stars can be resolved in M31, so individual bright stars can be applied as extinction tracers to determine the extinction law in M31, such as the LMC and SMC.
In recent years, works on the dust extinction law in M31 have been gradually conducted.
\citet{1996ApJ...471..203B} derived the UV extinction law with the spectra of stars in OB association of M31 and determined that the extinction law shows an MW-type extinction curve but with a weaker 2175 $\,{\rm \AA}$ bump.
The extinction law obtained by \citet{2014ApJ...785..136D} with 5 dusty clumps located around the central region of M31 shows an obvious 2175 $\,{\rm \AA}$ bump with $R_V \approx 2.4 - 2.5$, indicating small grains near the center region of M31.

\citet{2015ApJ...815...14C} obtained the UV spectra of four significantly reddened supergiants located on the disk of M31 and constructed the model SEDs with a combination of a stellar model atmosphere and the parameterized extinction curves proposed by \citet[hereafter FM90]{1990ApJS...72..163F}.
By fitting the model SEDs to the observed spectra, they found different extinction curves towards different sightlines.
One supergiant J004412.17+413324.2 of the four supergiants in \citet{2015ApJ...815...14C} is the closest to the M31 bulge, and its extinction curve resembles the LMC 30-Dor extinction curve and an $R_V \approx 2$ CCM MW extinction curve, tallying with the extinction law near the center region of M31 derived by \citet{2014ApJ...785..136D}.
Another tracer J003944.71+402056.2 with $R_V \approx 3.3$ shows a similarity to the MW and the LMC average extinction curves.
The extinction curves of the other two traces of J004034.61+404326.1 and J003958.22+402329.0 look the same as $R_V \approx 2.5$ and are steeper than the average MW extinction curve.
The average extinction curves derived by \citet{2015ApJ...815...14C} along with those derived by \citet{2014ApJ...785..136D} and \citet{1996ApJ...471..203B} are also presented in Figure \ref{fig:M31_intro}.

\citet{2014ApJ...780..172D} maps the dust mass surface density in M31 by using images from the Spitzer Space Telescope and Herschel Space Observation, from which a dust extinction map can be calculated based on the dust emission.
\citet{2015ApJ...814....3D} also mapped the distribution of dust in M31 with a 25-pc resolution.
They derived that the median value of $A_V$ is 1 mag, with very little surface area having an $A_V > 3$ mag.
There is morphological agreement between the $A_V$ map derived by \citet{2015ApJ...814....3D} and that calculated from the dust mass surface density in \citet{2014ApJ...780..172D}.
However, \citet{2015ApJ...814....3D} suggests that the extinction map inferred from dust emission by \citet{2014ApJ...780..172D} may overpredict the observed extinction by a factor of $\approx 2.5$.
In addition, \citet{2016MNRAS.459.2262D} mapped the dust distribution in the central 180 arcsec region of the M31 bulge from near-UV to near-IR at a 2 pc resolution.
This high-resolution dust map helps with measuring the fraction of obscured starlight across the field.

Given the above, the few calculated extinction curves in M31 agree with each other and resemble those of the MW and LMC.
Because of the large scale and the complex interstellar environment, dust extinction curves towards more different sightlines in M31 are still needed.
In this work, samples of bright O-type and B-type supergiants in M31 from the Local Group Galaxies Survey (LGGS, \citealt{2016AJ....152...62M}) are selected as extinction tracers.
We improve the method adopted in \citet{2015ApJ...815...14C} by substituting the model extinction curves directly from the dust model for the FM90 parameterized extinction law.
After fitting the model SEDs to the observed SEDs, extinction curves towards more sightlines from UV to near-IR in M31 are derived.
Section \ref{sec:data} presents the observed data in detail.
The improved method we adopt is described in Section \ref{sec:method}.
Section \ref{sec:re} shows the results and discussions.
Our conclusions are finally summarized in Section \ref{sec:conclusion}.


\section{Data and Sample} \label{sec:data}

Early-type stars are usually free of circumstellar dust \citep{2018MNRAS.478.3467S} and have an enormous intrinsic brightness \citep{2019ApJS..241...32L}.
They are commonly used to probe the dust extinction law in the MW and external galaxies \citep{2018ARA&A..56..673G}.
For instance, because of the exceptional brightness of the galactic B3-5 hypergiant Cygnus OB2 \#12 \citep{2012A&A...541A.145C,2013ARep...57..527C}, it is a popular target for studying interstellar phenomena including interstellar extinction \citep{1973PASP...85...87C,2015MNRAS.449..741W,2016MNRAS.458..491M,2016BaltA..25...42M}.
In addition, \citet{2015ApJ...815...14C} adopted four reddened O-type and B-type supergiants to derive the dust extinction curves in M31.
It follows that early-type stars are considered to be typical extinction tracers.

In this work, O-type and B-type supergiants from LGGS are selected to construct a sample of extinction tracers.
By gathering photometric data from the LGGS, the United Kingdom Infrared Telescope (UKIRT, \citealt{2013ASSP...37..229I}), the Panoramic Survey Telescope and Rapid Response System release 1 (PS1) Survey \citep{2016arXiv161205560C}, the Panchromatic Hubble Andromeda Treasury (PHAT) Survey \citep{2014ApJS..215....9W}, the XMM-Newton Serendipitous Ultraviolet Source Survey (XMM-SUSS, \citealt{2019yCat.2356....0P}) and the Swift Ultraviolet and Optical Telescope (Swift/UVOT, \citealt{2005SSRv..120...95R}), the SEDs from UV to near-IR bands are constructed for all the tracers.

As the extinction law in the external galaxy is calculated, the foreground MW dust extinction must be taken into consideration for each photometric point.
We adopt $E(B-V) \approx 0.06$ mag \citep{1998ApJ...500..525S,2011ApJ...737..103S,2012AJ....144..142B,2020ApJ...905L..20R}, assuming an $R_V = 3.1$ CCM extinction curve as described in \citet{2015ApJ...815...14C} to remove the foreground extinction of the MW rather than the previous estimate $E(B-V) \approx 0.08$ mag used in \citet{1996ApJ...471..203B}.

The photometric data applied in this work and the selection criteria are described in the following subsections.

\subsection{LGGS}

The LGGS provides $UBVRI$ plus the interference-image photometry of luminous stars in spiral galaxies M31 and M33 using the Kitt Peak National Observatory 4m telescope, along with those found in seven dwarf systems currently forming massive stars (IC 10, NGC 6288, WLM, Sextans A and B, Pegasus, and Phoenix) \citep{2006AJ....131.2478M,2007AJ....134.2474M,2007AJ....133.2393M,2011AJ....141...28M}.
For	M31, the catalog contains 371,781 stars, of which 64 and 321 are respectively confirmed as O-type and B-type stars by the 6.5m MMT telescope with the 300 fiber positioner Hectospec \citep{2016AJ....152...62M}.

\begin{figure}[ht!]
	\includegraphics[scale=0.1]{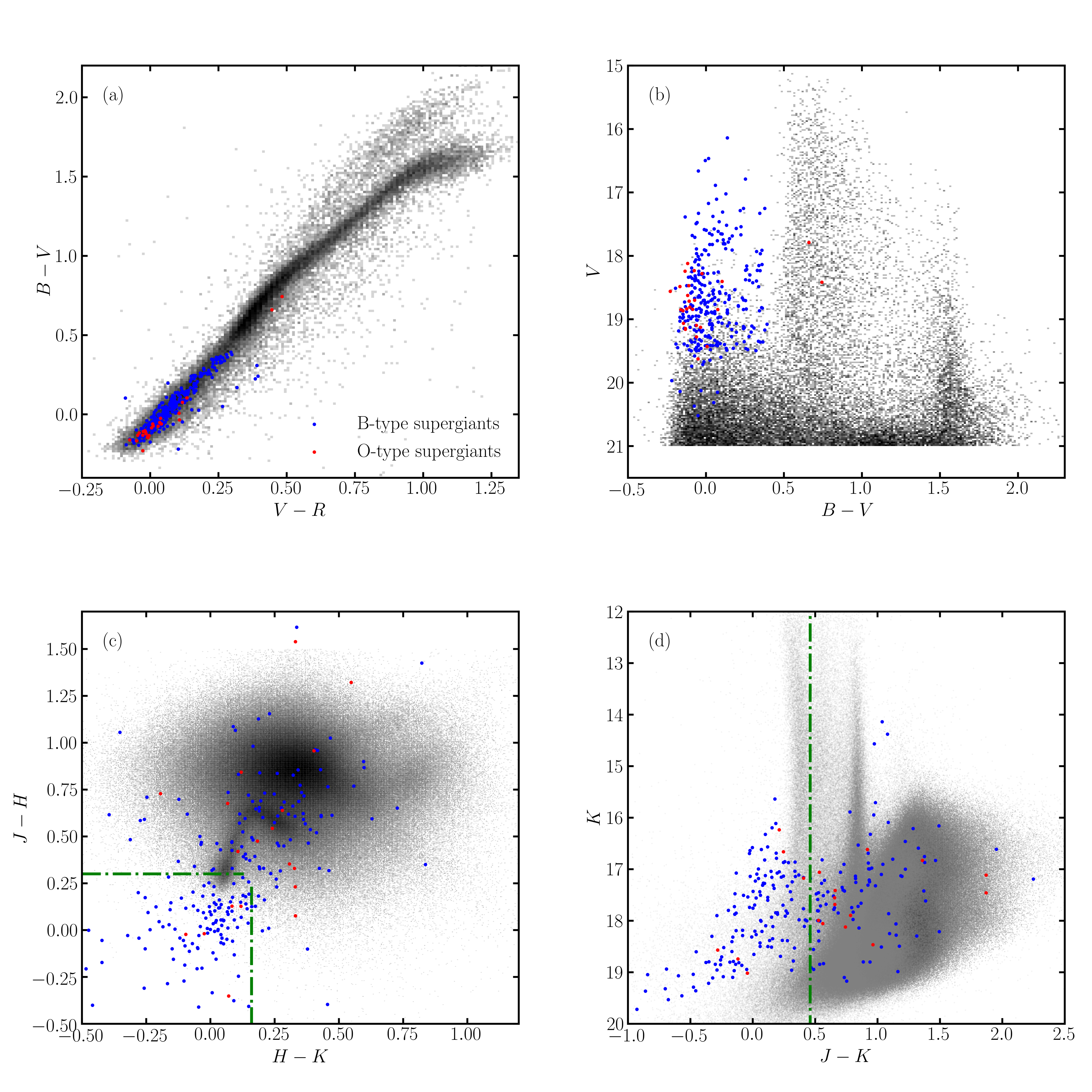}
	\caption{Color-color diagrams (CCDs) and color-magnitude diagrams (CMDs) for all LGGS or UKIRT sources (background gray dots) and the selected supergiants as extinction tracers.
	In panel (a) and panel (b), the red dots indicate all 27 O-type supergiants in the sample, while the blue dots are all 281 B-type supergiants in the sample.
	In panel (c) and panel (d), the red dots and the blue dots are the O-type supergiants and B-type supergiants with near-IR photometry in the sample.
	The green dashed-and-dotted line indicates the criteria for emission in near-IR bands, as described in Section \ref{subsec:UKIRT}.
	Near-IR photometry of sources within the green dashed-and-dotted line in panel (c) and on the left side of the green dashed-and-dotted line in panel (d) is adopted.
	\label{fig:samples} }

\end{figure}

According to \citet{2016AJ....152...62M}, all the O-type and B-type supergiants are considered M31 members.
Thus, the O-type and B-type isolated supergiants in the LGGS catalog\footnote{The column named `Cwd' in the LGGS catalog shows the index for the degree of crowding for each star.
Cwd = `I' indicates that stars isolated with contamination $< 5\%$ \citep{2016AJ....152...62M}.}
are selected \citep{2016AJ....152...62M} to construct the extinction sample. However, due to the limitation of the resolution of ground-based telescopes, OB associations or binaries may be identified as single OB stars. Therefore, we use PHAT/F475W images to check the reliability of the extinction sample and eliminate the suspect ones before the calculation. 
We also set the upper limit of the $V$ band magnitude to 21 mag to ensure a decent signal-to-noise ratio (SNR) \citep{2006AJ....131.2478M}.
The extinction sample contains 27 O-type supergiants and 281 B-type supergiants in total.
Figure \ref{fig:samples} (a) and (b) show the $V-R/B-V$ diagram and $B-V/V$ diagram for all LGGS sources and the selected supergiants.

\subsection{UKIRT/WFCAM} \label{subsec:UKIRT}

The Wide Field Camera (WFCAM) on the 3.8m UKIRT images the sky in three near-IR broadband filters, $J, H, K$, centering at 1.25, 1.63 and 2.20 $\mu {\rm m}$, respectively \citep{2008A&A...487..131C,2013ASSP...37..229I,2020ApJ...889...44N}.
The $JHK$ brightness of 27 O-type supergiants and 263 B-type supergiants in the sample can be found in the UKIRT catalog processed by \citet{2021ApJ...907...18R}.

Excess emission above the extrapolated stellar continuum is often detected in the near-IR range \citep{1984MNRAS.209..111J,2018ARA&A..56..673G}, probably due to nebular emission \citep{2009AJ....138..130S}, hot equilibrium dust \citep{2000A&A...363..493V} or small, out-of-equilibrium grains \citep{2010AJ....140.2124B}.
In this work, we calculate the IR excesses by estimating the maximum values of $(J-H)$ and $(H-K)$: 
\begin{equation}
    (J-H) = (J-H)_0 + E(J-H),
\end{equation}
\begin{equation}
	(H-K) = (H-K)_0 + E(H-K).	
\end{equation}
For the part of intrinsic color index, according to Table 7.8 in \citet{2000asqu.book.....C}, the intrinsic color index of $(J-H)_0$ for O-type and B-type supergiants is usually smaller than 0.09 mag, while that of $(H-K)_0$ is no larger than 0.03 mag.
These values are larger than those give in \citet{1994MNRAS.270..229W} and \citet{2006A&A...457..637M}.
In addition, $(J-H)_0 = 0.09$ mag is also larger than the bluest $(J-H)$ in each sub-class for the 646 O-type and B-type supergiants obtained by \citet{2020AJ....159..208D}.
In order to keep more reasonable near-IR data for the calculation, $(J-H)_0 = 0.09$ mag and $(H-K)_0 = 0.03$ mag are adopted as the upper limits of intrinsic color index.
For the part of color excess,
the upper limit of $A_V$ is set to 2 mag based on the extinction of M31 applied to the model in \citet{2014ApJS..215....9W}.
$(\frac{A_J}{A_V}-\frac{A_H}{A_V})_{\rm max}$ and $(\frac{A_H}{A_V}-\frac{A_K}{A_V})_{\rm max}$ are derived by $A_{\lambda}/A_V = (0.372 \pm 0.003){\lambda}^{-2.07 \pm 0.03}$ proposed in \citet{2019ApJ...877..116W} by assuming the general interstellar environment in M31 resembles that of the MW.

In this way, we consider $(J-H) < 0.3$ mag and $(H-K) < 0.16$ mag as the selection criteria in order to eliminate the influence of near-IR emission, based on which near-IR photometry for 5 O-type supergiants and 112 B-type supergiants is adopted.
Figure \ref{fig:samples} (c) and (d) present the $H-K/J-H$ diagram and $J-H/K$ diagram for all UKIRT sources and the selected tracers.

\subsection{PS1 Survey}

Pan-STARRS \citep{2004AN....325..636H,2016arXiv161205560C} release 1(PS1) survey used a 1.8m telescope to perform images covering a wavelength range from 400 $nm$ to 1 $\mu {\rm m}$ in five broadband filters: $g, r, i, z, y$, with wavelengths centered at 0.481, 0.617, 0.752, 0.866 and 0.962 $\mu {\rm m}$, respectively \citep{2010ApJS..191..376S,2012ApJ...756..158S,2012ApJ...750...99T}.

27 O-type and 276 B-type supergiants in the sample can be found as counterparts in the PS1 catalog.
For these tracers, we first examine the information flag for each photometric point and select reliable photometric data with good qualities.
We then choose photometry with a magnitude error no greater than 0.1 mag.
As a result, the photometry of 17 O-type supergiants and 176 B-type supergiants is maintained.

\subsection{PHAT Survey}

The PHAT survey covers $\approx$ 1/3 of the star-forming disk of M31 in near UV ($\lambda_{F275W} = 0.272~\mu {\rm m}$, $\lambda_{F336W} = 0.336~\mu {\rm m}$), optical ($\lambda_{F475W} = 0.473~\mu {\rm m}$, $\lambda_{F814W} = 0.798~\mu {\rm m}$), and near IR ($\lambda_{F110W} = 1.120~\mu {\rm m}$, $\lambda_{F160W} = 1.528~\mu {\rm m}$) bands and contains 117 million equidistant stars, with very little ($\ll 1 \%$) contamination from the MW foreground or background galaxies \citep{2014ApJS..215....9W}.

14 O-type supergiants and 108 B-type supergiants in the sample are recorded with photometry in the PHAT catalog.
We check the star reliability GST flag listed in the catalog for each photometric point of each tracer and select the photometry with star reliability GST flag = `T'.
PHAT photometry for 4 O-type supergiants and 76 B-type supergiants can thus be applied.

\subsection{XMM-SUSS}

XMM-SUSS is a catalog of UV sources detected serendipitously by the XMM-Newton observatory \citep{2001A&A...365L..36M}, of which SUSS4.1 is a new 2018 release \citep{2019yCat.2356....0P}.
The AB magnitude distributions peak at $m_{AB} = 20.2, 20.9$ and 21.2 mag in $UVW2~(\lambda_{\rm eff} = 2120 $ \,{\rm \AA}), $UVM2~(\lambda_{\rm eff} = 2310 $ \,{\rm \AA}) and $UVW1~(\lambda_{\rm eff} = 2910 $ \,{\rm \AA}), respectively.

Cross-matching with XMM-SUSS 4.1 in 1$^{\prime \prime}$ provides UV photometry of 23 O-type and 227 B-type supergiants in our sample.
We checked the quality for each photometric point based on the `source quality flag' column presented in the catalog and keep the photometry with source quality flag = `FFFFFFFFFFFF'.
In addition, some photometry is so bright that it could not fit the whole SED well.
This kind of photometry is considered foreground photometry and is rejected in this work.
Photometric data of XMM SUSS 4.1 are only adopted for 6 O-type and 49 B-type supergiants in the sample.

\subsection{Swift/UVOT}

Swift/UVOT \citep{2005SSRv..120...95R} also provides photometry in three UV bands ($\lambda_{UVW2} = 0.209~\mu {\rm m}$, $\lambda_{UVM2} = 0.225~\mu {\rm m}$, $\lambda_{UVW1} = 0.268~\mu {\rm m}$), which is organized in the Swift/UVOT Serendipitous Source catalog \citep{2015yCat.2339....0Y}.

By cross-matching, 14 O-type and 118 B-type supergiants in the sample are found in the corresponding UV photometry.
The extended flag and quality flag for each photometric point in the catalog are first checked, and then photometry with the extended flag = `0' and quality flag = `0' is selected.
In addition, Swift/UVOT data that could not fit the entire SED well were considered foreground data and were eliminated.
Finally, UVOT photometry for 6 O-type and 53 B-type supergiants in the sample is retained.

~\\
Above all, at most 25 bands of the photometric data from UV to near-IR can be obtained for each star.
Table \ref{Tab:data} summarizes the selection criteria and the number of photometry point adopted for each catalog.
We introduce \emph{pcFlag} to show the coverage of passbands adopted in the calculation.
\emph{U} in \emph{pcFlag} means the results are derived with UV data (here refers to the passbands bluer than $U$ band), while \emph{V} and \emph{I} in \emph{pcFlag} are short for visual bands and near-IR bands (here refers to the passbands redder than $y$ band), respectively.
In this work, the extinction law in M31 is discussed mainly based on the results of the sightlines with \emph{pcFlag = `UVI'}.

\begin{deluxetable*}{cccccccccc}
	\tablecaption{Photometry sources, selection criteria and number of photometries adopted for each catalog \label{tab:data}}
		\tablehead{	
		\colhead{Catalog} & \colhead{Selection criteria} &\colhead{O-type supergiants} &  \colhead{B-type supergiants}   \label{Tab:data}
		}
	\startdata
		LGGS$^a$     & I. Supergiants  & 27 & 281 \\
		             & II. Cwd = `I' & & \\
					 & III. V $<$ 21 mag & & \\
		\hline
		UKIRT/WFCAM  & I. $(J-H) < 0.3$ mag & 5  & 112  \\
		             & II. $(H-K) < 0.16$ mag & & \\
		PS1 Survey   & I. Check information flags & 17 & 176 \\
		             & II. Magnitude error $<$ 0.1 mag & & \\
		PHAT Survey  & I. Star reliability GST flag = `T' & 4  & 76  \\
		XMM-SUSS     & I. Source quality flag = `FFFFFFFFFFFF' & 6  & 49  \\
		             & II. Eliminate foreground photometry & & \\
		Swift/UVOT   & I. Extended flag = `0' & 6  & 53  \\
		             & II. Quality flag = `0' & & \\
		             & III. Eliminate foreground photometry & & \\
	\enddata
\tablecomments{$^a$ The LGGS catalog contains 64 O-type and 321 B-type stars for M31.
The selected 27 O-type and 281 B-type isolated supergiants constitute the sample of extinction tracers in this work. }
\end{deluxetable*}


\section{Method} \label{sec:method}

First used by \citet{1970IAUS...36...28B}, the ``pair method'' is extensively adopted to determine the interstellar extinction curve, which compares the spectrum of a reddened star with that of an unreddened star with the same or similar spectral type.
Based on the traditional pair method, an improved method is proposed in this work that forward models the SEDs with a combination of the intrinsic spectra from the stellar model atmosphere and the extinction curves derived from the dust model.
After fitting the model SEDs to the observed data, the extinction law can be derived, the dust properties can be further analyzed based on the derived parameters in the dust model, and even the stellar parameters can be obtained.

\begin{figure}
	\centering
	\includegraphics[scale=0.8]{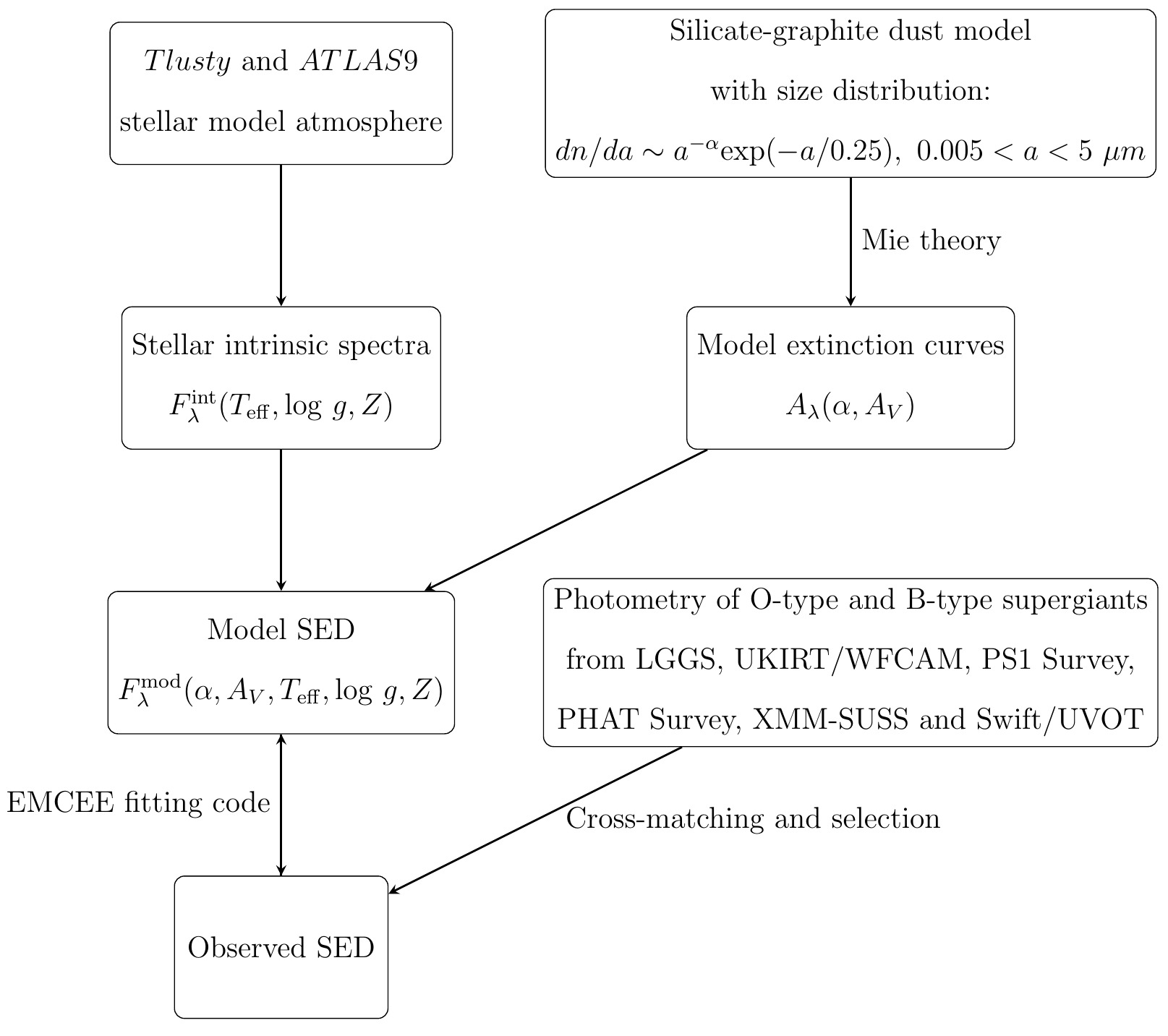}
	\caption{The improved pair method applied in this work.
	The details of the observed data are described in Section \ref{sec:data}.
	Intrinsic spectra from the model atmosphere and the model extinction curves derived from the dust model are introduced in Section \ref{subsec:intrinsic_sed} and Section \ref{subsec:ext}, respectively.
	The construction of the model SEDs is presented in Section \ref{subsec:fullsed}.
	Section \ref{subsed:com} shows the fitting technique.
	\label{fig:method}}
\end{figure}

Figure \ref{fig:method} summarizes the improved pair method used in this work, and the following subsections show the details.

\subsection{Intrinsic SEDs} \label{subsec:intrinsic_sed}

The traditional pair method considers the spectrum of an unreddened star with a spectral type similar to that of the extinction tracer as the intrinsic spectrum.
However, there are not enough comparison stars because of the limited sample of unreddened or slightly reddened stars in M31.
We thus adopt stellar atmosphere models as ``pair stars" to obtain intrinsic spectra, which are as accurate as using the spectra of real stars \citep{2005AJ....130.1127F}.

A Tlusty stellar model atmosphere \citep{2017arXiv170601859H,2017arXiv170601937H,2017arXiv170601935H} and an ATLAS9 stellar model atmosphere \citep{2003IAUS..210P.A20C} are applied in this work to obtain the intrinsic spectra.
The Tlusty stellar model atmosphere was first proposed in \citet{1988CoPhC..52..103H}, and with specific stellar parameters and atomic and ionic energy levels given, it can be used to calculate plane-parallel, horizontally homogeneous model stellar atmospheres in hydrostatic and radiative (or radiative + convective) equilibrium to derive the intrinsic spectra.
We adopt the two extensive grids of a non-local thermodynamic equilibrium (NLTE) fully metal-line blanketed model atmosphere for O-stars \citep{2003ApJS..146..417L} and early B-stars \citep{2007ApJS..169...83L}.
For the O-type stars, the grid considers 12 values of effective temperatures ($27000~{\rm K} \leq T_{\rm eff} \leq 55000~{\rm K}$ with 2500 K steps), 8 surface gravities [$3.0 \leq {\rm log}(g) \leq 4.75$ with 0.25 dex steps] and 10 chemical compositions.
For B-type stars, there are 16 values of the effective temperatures ($15000~{\rm K} \leq T_{\rm eff} \leq 30000~{\rm K}$ with 1000 K steps), 13 values of the surface gravities [$1.75 \leq {\rm log}(g) \leq 4.75$ with 0.25 dex steps] and 6 chemical compositions in the grid.
The ATLAS codes written by Kurucz are quite popular \citep{1970SAOSR.309.....K} among codes of computing LTE blanketed stellar atmosphere model.
The ATLAS9 grid \citep{2003IAUS..210P.A20C} considers effective temperature $T_{\rm eff}$ from 3500 K to 50000K, surface gravity log($g$) from 0.0 to 5.0 and several chemical compositions. 

We combine the non-LTE Tlusty OSTAR and BSTAR girds \citep{2003ApJS..146..417L,2007ApJS..169...83L} and the LTE ATLAS9 grid \citep{2003IAUS..210P.A20C} to construct a series of model intrinsic spectra.
Both grids derive very similar spectra for the same stellar parameters in the overlap region.
In this work, the Tlusty grid is preferred over the ATLAS9 grid in the regions of overlap as mentioned in \citet{2016ApJ...826..104G}.



\subsection{Dust Extinction Model} \label{subsec:ext}

Instead of parameterized extinction curves such as the CCM \citep{1989ApJ...345..245C} and FM90 \citep{1990ApJS...72..163F} extinction laws widely used in other works, extinction curves directly derived from the dust model are applied in this work.
Along the sightline, the extinction at the wavelength of $\lambda$ can be expressed as:
\begin{equation} \label{eq:3}
    A_{\lambda}/N_{\rm H} = 1.086 \int_{a_{\rm min}}^{a_{\rm max}} C_{\rm ext}(a, \lambda) \frac{1}{n_{\rm H}} \frac{dn}{da} da,
\end{equation}
where $a$ is the radius of the dust grain, which is assumed to be spherical, $C_{\rm ext}(a,\lambda)$ is the extinction cross section and can be calculated by Mie theory \citep{1984ApJ...285...89D}, $n_{\rm H}$ is the number density of H nuclei, $N_{\rm H}$ is the column density of H nuclei, and $dn/da$ is the dust size distribution.
A classic silicate-graphite dust model is used in this work, which consists of amorphous silicate and graphite.

\citet{1994ApJ...420...87Z} determined the metallicity in M31 to be approximately twice the solar abundance, and it was also suggested to be close to the solar value by \citet{2000ApJ...541..610V} and \citet{2001MNRAS.325..257S}.
A recent study by \citet{2012ApJ...758..133S} reconfirmed the supersolar abundances in M31.
\citet{2014ApJ...780..172D} estimated the interstellar medium metallicity to vary from $Z/Z_{\bigodot} \approx 3$ at $R=0$ to $Z/Z_{\bigodot} \approx 0.3$ at $R = 25$ kpc.
However, detailed information on the abundance of M31 is still uncertain.
The solar abundances have historically been taken to represent the total interstellar abundances \citep{2005ApJ...622..965L}.
The interstellar abundances in M31 can thus be assumed to be similar to those in the MW with protosolar values \citep{2009ARA&A..47..481A}.
We consider the mass ratio of graphite to silicate to be $f_{cs} = 0.3$ as a typical value used in previous works \citep{2013EP&S...65.1127G,2014P&SS..100...32W,2015ApJ...807L..26G,2020P&SS..18304627G} and adopted in \citet{2014ApJ...780..172D}, which means that the elements of Fe, Mg and Si are all in the solid phase and constrained in silicate dust, and that the fraction of gas-phase carbon is 50\%.

Silicate-graphite dust models with a power-law size distribution ($dn/da \sim a^{-\alpha}$) and an exponential cutoff power-law size distribution [$dn/da \sim a^{-\alpha}{\rm exp}(-a/a_c)$] are popular for investigating the dust extinction law.
The power-law size distribution was proposed by Mathis, Rumpl and Nordsieck (hereafter MRN, \citealt{1977ApJ...217..425M}).
They suggested the expression $dn/da \sim a^{-3.5}$, extending from a small grain ($a_{\rm min} = 0.005~\mu {\rm m}$) to a large ($a_{\rm max} = 0.25~\mu {\rm m}$) size cutoff\footnote{
The size distributions found by MRN are roughly power laws with an exponent of approximately -3.3 to -3.6.
The size range for graphite is from $\approx$ 0.005 $\mu {\rm m}$ to $\approx$ 1 $\mu {\rm m}$.
Other materials (silicate) are distributed in a narrower range: $\approx 0.025-0.25~\mu {\rm m}$.
For convenience, this MRN dust size distribution is commonly considered to be a power law $dn/da \sim a^{-3.5}$, $0.005 < a < 0.25~\mu {\rm m}$ for a mixture of silicate and graphite.
}.
Kim, Martin and Hendry (hereafter KMH, \citealt{1994ApJ...422..164K}) added an exponential cutoff into the simple power law, changing the expression to $dn/da \sim a^{-\alpha}{\rm exp}(-a/a_c)$ with the dust size distribution curve descending when dust size is larger than the exponential cutoff size $a_c$.

Considering the sudden cutoff of the dust size in the MRN model and the degeneracy of two parameters in the KMH model, here we fix the value of $a_c$ in the KMH model to 0.25 $\mu {\rm m}$, which is the maximum cutoff value of dust size in the MRN model, and introduce a mixed dust size distribution in this work:
\begin{equation}
dn/da \sim a^{-\alpha}{\rm exp}(-a/0.25), a_{\rm min} < a < a_{\rm max}.
\end{equation}
The lower and upper values of the dust radius are set to $a_{\rm min} = 0.005~\mu {\rm m}$, and $a_{\rm max} = 5~\mu {\rm m}$, respectively.

\subsection{Full SED Model} \label{subsec:fullsed}
The observed SED $F^{\rm obs}_{\lambda}$ of a reddened star can be expressed as \citep{2005AJ....130.1127F}:
\begin{equation}
F^{\rm obs}_{\lambda}=F^{\rm int}_{\lambda}{\theta}^2 10^{-0.4A_{\lambda}},
\end{equation}
where $F^{\rm int}_{\lambda}$ is the intrinsic surface flux of the star at wavelength ${\lambda}$, ${\theta} \equiv (R/d)^2$ is the angular radius of the star (where $d$ is the distance and $R$ is the stellar radius), and $A_{\lambda}$ is the absolute attenuation of the stellar flux by intervening dust at $\lambda$.
With the combination of the intrinsic spectra and the model extinction curves mentioned in the previous two subsections, the model monochromatic flux extinguished by dust can be written as:
\begin{equation}
F_{\lambda}^{\rm mod}(\alpha, A_V, T_{\rm eff}, {\rm log}~g, Z)=F^{\rm int}_{\lambda}(T_{\rm eff}, {\rm log}~g, Z){\theta}^2 10^{-0.4A_{\lambda}(\alpha,A_V)},
\end{equation}
where the effective temperature $T_{\rm eff}$, the surface gravity ${\rm log}(g)$, and the metallicity $Z$ parameterize the stellar model atmosphere.
$A_V$ is the extinction in magnitudes in the $V$ band, and $\alpha$ defines the dust size distribution in the dust model and the shape of the extinction curve.

To compare with the photometry, the model band flux can be calculated as:
\begin{equation}
F^{\rm mod}_i = \frac{\int \lambda G_i(\lambda)F_{\lambda}^{\rm mod}d \lambda}{\int \lambda G_i(\lambda)d \lambda}
\end{equation}
where $G_i({\lambda})$ is the bandpass response function for the $i_{\rm th}$ band.
The flux for each band is thus obtained from the response function and model spectra with the intrinsic spectra extinguished by interstellar dust.


\subsection{Fitting Technique} \label{subsed:com}

Markov Chain Monte Carlo (MCMC) analyse is widely used for sampling approximation and posterior probability distribution functions (PDFs).
Using this method, dust extinction maps in the MW can be derived \citep{2014MNRAS.443.2907S, 2019ApJ...887...93G, 2021ApJ...906...47G}.
In addition, \citet{2016ApJ...826..104G} took advantage of the MCMC method to propose a Bayesian Extinction And Stellar Tool (BEAST) for modeling the dust extinguished SED.
Moreover, it helped prompt several other extinction works on external galaxies \citep{2015ApJ...815...14C,2020ApJ...888...22V}.
In this work, this MCMC approach based on Bayesian theory is applied to fit the model SEDs to the observed SEDs.

In Bayesian probability theory \citep{1946AmJPh..14....1C}, posterior PDF for the parameter is crucial in the probabilistic data analysis procedure.
The Gaussian likelihood is applied and flat prior is imposed on $\alpha,~{\rm log}(g)$ and $A_V$ in this work.
The fitting parameter $\alpha$ is set to be 0.5 - 6.0 with a 0.1 resolution, while $A_V$ is set to be 0 - 5 mag with a resolution of 0.1 mag.
Regarding the effective temperature log$(T_{\rm eff})$ and gravity log$(g)$, the priors are based on the spectral type and the Tlusty and the ATLAS9 stellar model atmosphere grids mentioned in Section \ref{subsec:intrinsic_sed}.
The calibration of spectral type to effective temperature log$(T_{\rm eff})$ and surface gravity log$(g)$ refers to \citet{2000asqu.book.....C} and \citet{2008flhs.book.....C}.
Gaussian prior is imposed on log($T_{\rm eff}$) based on the spectral type given in \citet{2016AJ....152...62M} with one subclass in spectral type considered as uncertainty \citep{2015ApJ...815...14C}.
For the metallicity log$(Z)$, as described in Section \ref{subsec:ext}, we first adopt the solar metallicity and then twice the solar metallicity, and find that these two values lead to similar results.
Therefore, only the solar metallicity for M31 is adopted in this work.

As a result, a comprehensive grid containing hundreds of thousand sets of model SEDs is constructed.
The EMCEE \citep{2013PASP..125..306F} fitting code is used in this work to fit our model SEDs to the observed data.
With this MCMC ensemble sampler, we can find the most suitable parameters and the corresponding confidence intervals.

\subsection{Results Screening} \label{subsec:selection}

\begin{figure}
	\centering
	\includegraphics[scale=0.14]{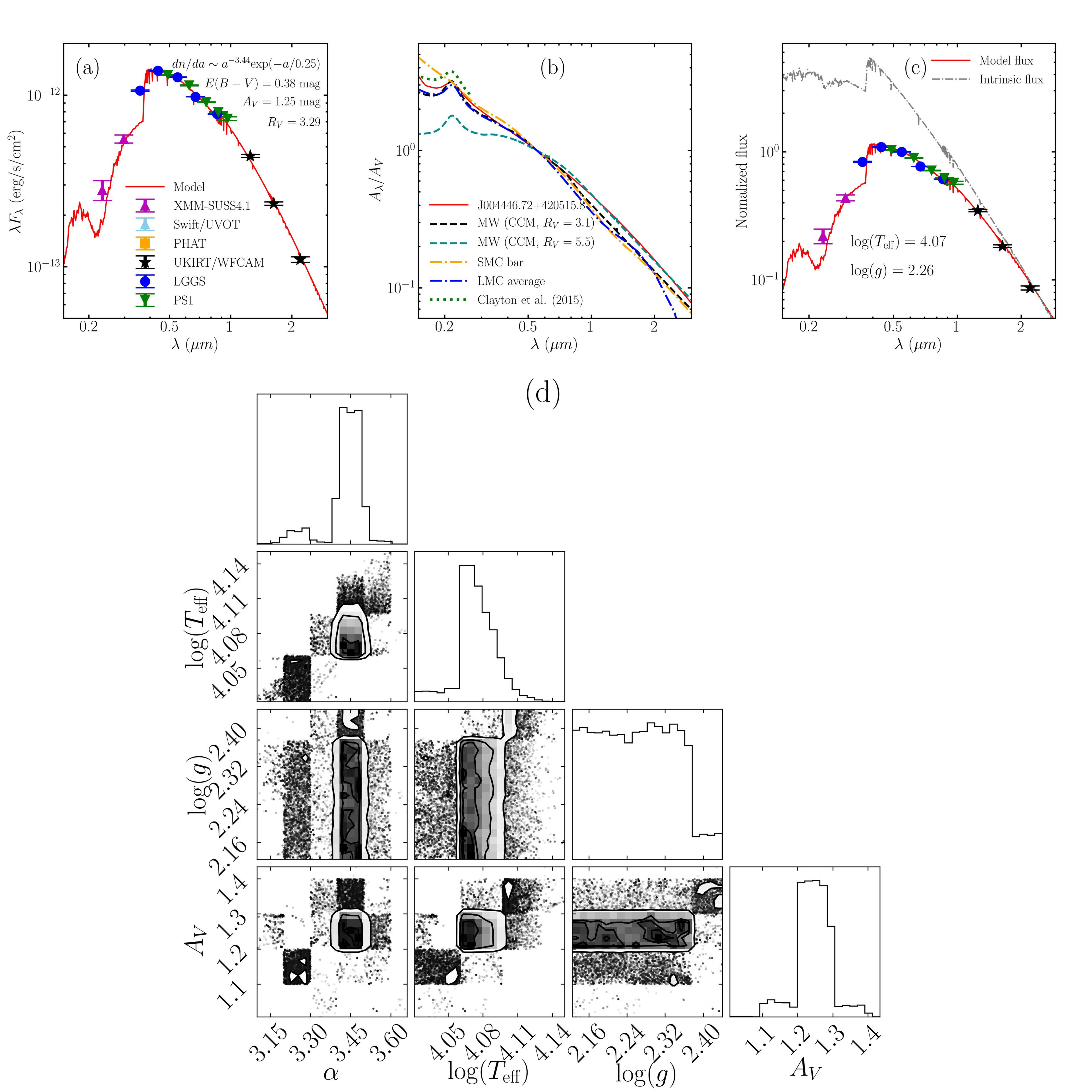}
	\caption{Example results for a B-type supergiant J004446.72+420515.8 in the sample.
	(a): The best-fitting model SED compared with the observed data.
	The fitting parameters $\alpha$ and $A_V$ and the corresponding $R_V$ and $E(B-V)$ are presented.
	(b): The extinction curve derived from the dust model compared to the average extinction curve in \citet{2015ApJ...815...14C} and those of the average LMC, the bar region in the SMC, and the diffuse region in the MW.
    (c) Normalized intrinsic spectrum and best-fitting SED to the observed data with the effective temperature log$(T_{\rm eff})$ and the gravity log$(g)$ marked.
	(d) One- and two-dimensional projections of the posterior probability distributions of the derived parameter [$\alpha,~{\rm log}(T_{\rm eff}),~{\rm log}(g),~A_V$].
	\label{fig:example}}
\end{figure}

With the model SEDs fitting to the observed data, the fitting parameters, including $\alpha$ in the dust model, extinction in the $V$ band $A_V$, the effective temperature log($T_{\rm eff}$) and the gravity ${\log}(g)$ for each tracer, can be derived, as well as the corresponding extinction curve, the color excess $E(B-V)$, and the total-to-selective extinction ratio $R_V$.
Figure \ref{fig:example} presents an example result of a B-type supergiant named J004446.72+420515.8.
Panel (a) shows the best-fitting model SED compared to the observed SED with $\alpha$, $A_V$, $R_V$ and $E(B-V)$ marked.
Panel (b) presents the extinction curve towards the sightline of the tracer with those in the MW and MCs, as well as the average extinction curve derived by \citet{2015ApJ...815...14C}.
Panel (c) compares the normalized intrinsic spectrum and the model SEDs with the observed data, in which the derived parameters log$(T_{\rm eff})$ and log$(g)$ are also labeled.
Panel(d) is divided into 16 subgraphs.
The four subgraphs located on the diagonal from upper left to lower right are the posterior probability distributions of $\alpha$, log$(T_{\rm eff})$, log$(g)$ and $A_V$ generated from the EMCEE results.
The nearly uniform distribution of log$(g)$ is due to the insensitivity of the intrinsic spectra of O-type and B-type supergiants to the surface gravity.
Other subgraphs show the two-dimensional projections of the posterior probability distributions of the corresponding parameters marked on the left side and below.

As the results of all tracers presented, tracers with unreliable and unreasonable results should be eliminated for further analysis.
We keep the tracers as follows:

I. The fraction of proposed steps that are accepted in the MCMC process, acceptance fraction, is in the range of 0.2-0.5 \citep{2013PASP..125..306F}, which indicates a good MCMC performance.

II. The derived color excess $E(B-V) > 0.16$ mag is considered to be a reliable result.
Slightly reddened stars [$E(B-V) \approx 0.1$ mag] may lead to larger errors.
In \citet{2015ApJ...815...14C}, extinction curves for stars with $E(B-V)$ less than 0.16 mag were not used in the analysis.

III. The results with the derived total-to-selective extinction ratio in the range of $R_V =1.5-7$ are retained.
Observationally, $R_V$ can be as small as $\approx 2$ towards some diffuse sightlines \citep{1992ApJ...393..193W,1999PASP..111...63F,2011piim.book.....D,2017ApJ...848..106W} and as large as $\approx 6$ in dense molecular clouds \citep{1990ARA&A..28...37M,1999PASP..111...63F,2011piim.book.....D}.
Values of $R_V$ beyond the range of $1.5-7$ are taken to be unphysical.
More physical $R_V$ values and more reliable extinction results can be expected, provided that the UV data are adequate (see Section \ref{subsec:rv} for details).

Based on the selection criteria mentioned above, 9 O-type and 137 B-type supergiants in the sample are selected with reliable and reasonable results for further analysis.
In the selected sample there are 105 tracers with near-IR data (near-IR in this work refers to passbands with effective wavelengths longer than $y$: $J,~H,~K,~F110W,~F160W$), and 79 tracers with UV data (UV in this work refers to passbands with effective wavelengths shorter than $U$: $UVW2,~UVM2,~UVW1,~F275W$).


\section{Result and Discussion} \label{sec:re}

\subsection{The Extinction Curves of M31} \label{subsec:eachre}

The results for each tracer with the ID and spectral type are partly listed in Table \ref{tab:re}.
The ID and spectral type are obtained from the LGGS catalog \citep{2016AJ....152...62M}.
The fitting parameters $\alpha$, log$(T_{\rm eff})$, log$(g)$ and $A_V$ and their uncertainties are tabulated based on the 50\%, 16\% and 84\% values of the marginalized 1D posterior probability distribution functions generated from the EMCEE results.
The seventh and eighth columns in Table \ref{tab:re} are the corresponding derived $E(B-V)$ and $R_V$, respectively.
$E(B-V)$ is calculated by $E(B-V) = A_V(A_B/A_V-1)$, and then $R_V = A_V/E(B-V)$.
The ninth and tenth columns in Table \ref{tab:re} present the number of passbands and the coverage of the passbands used in the calculation (\emph{pcFlag} mentioned in Section \ref{sec:data}).
The columns following the \emph{pcFlag} column are the photometry in each band, which are available in a machine-readable format.

\begin{figure}
	\centering
	\includegraphics[scale=0.9]{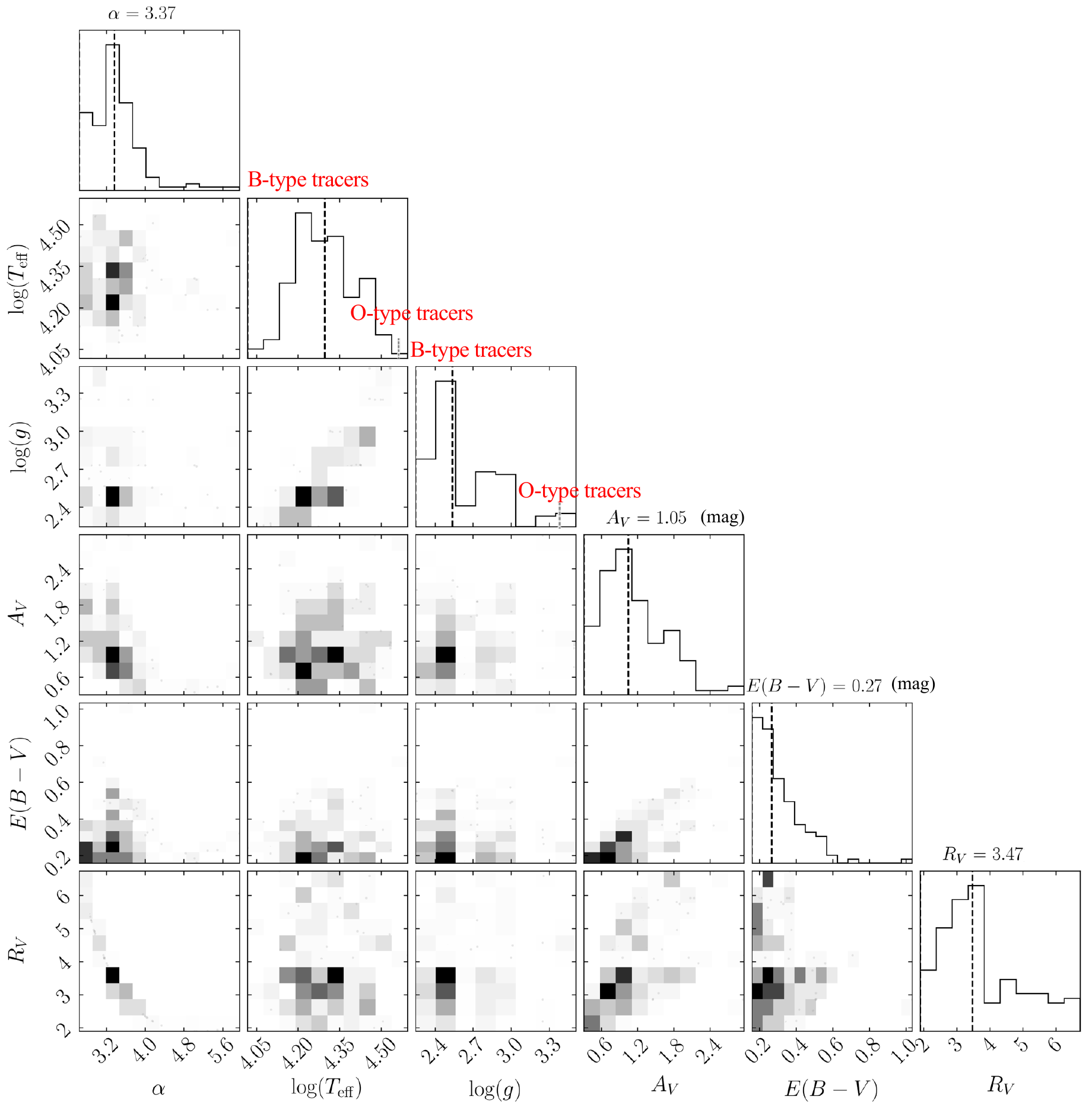}
	\caption{Distributions and correlations of the expectation values of the four fitting parameters [$\alpha,{\rm log}(T_{\rm eff}), {\log}(g), A_V$] and the two derived parameters [$E(B-V)$, $R_V$] for the selected tracers in the extinction sample.
    The two peaks in the ${\rm log}(g)$ distribution indicate the O-type and B-type tracers, respectively.
	The relatively inapparent peaks for the O-type tracers are due to the small number.
	\label{fig:pm_dis}}
\end{figure}

Figure \ref{fig:pm_dis} shows the distributions of the expectation values of the six parameters [$\alpha, {\log}(T_{\rm eff}), {\rm log}(g), A_V$, $E(B-V)$, $R_V$] for each selected tracer and the correlations between the parameters.
The median values of the parameters are also presented: $\alpha = 3.37$, $A_V = 1.05$ mag, $E(B-V) = 0.27$ mag, $R_V = 3.47$ for all selected tracers.
There seems to be no correlations or degeneracies between $\alpha$, $A_V$, log$(T_{\rm eff})$ and log$(g)$ that could cause problems in this work.
The usually seen degeneracy between log$(T_{\rm eff})$ and $A_V$ will be further discussed in Section \ref{subsec:sensi_test}.


Table \ref{tab:re_dis} summarizes the median results with the upper and lower limits for each parameter, which are extracted based on all the derived values of each parameter for all the selected tracers.
Lines 1 to 4 in Table \ref{tab:re_dis} present the results of the four subsamples based on the \emph{pcFlag} mentioned above,and the fifth line shows the results for all the selected tracers in the extinction sample.
In this work, the general extinction law in M31 refers to the results of the sightlines with \emph{pcFlag = `UVI'}.
The sixth and seventh lines are the results for the 9 O-type and 131 B-type selected supergiants in the sample, respectively.
Lines 8 to 11 in Table \ref{tab:re_dis} present the influence of a lack of UV or near-IR on the results (see Section \ref{subsec:IRUV} for details).
The last line in Table \ref{tab:re_dis} shows the results of fitting our model extinction curves derived directly from the silicate-graphite dust model (see Section \ref{subsec:ext} for details) to that of the diffuse region in the MW (CCM extinction curve with $R_V = 3.1$) for a comparison.

\begin{figure}
	\centering
	\includegraphics[scale=0.8]{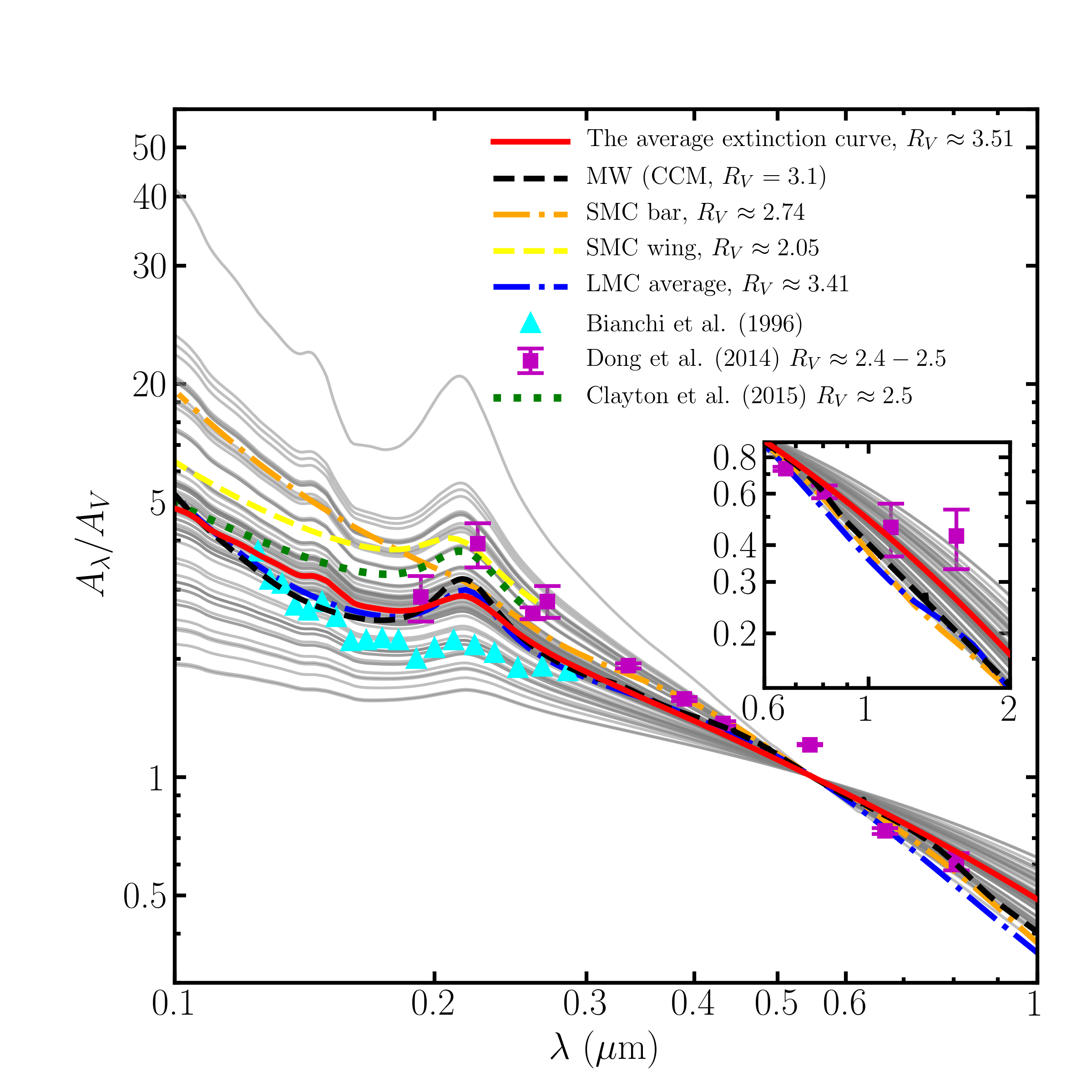}
	\caption{Extinction curves for the tracers (gray solid lines) in M31, of which the red solid line presents the average extinction curve.
	This shows the extinction laws from UV ($0.1~\mu {\rm m})$ to $1~\mu {\rm m}$, and the inset shows those from $0.6~\mu {\rm m}$ to near-IR bands ($2~\mu {\rm m}$).
	The black dashed line shows the extinction law of the diffuse region in the MW.
	The orange dashed-and-dotted line, the yellow dash line and the blue dashed-and-dotted line are the extinction curves of the SMC bar and the SMC wing and the average extinction curve of LMC, respectively \citep{2003ApJ...594..279G}.
	The cyan triangles show the flat extinction curve in M31 derived by \citet{1996ApJ...471..203B}.
	The purple squares indicate the extinction law around the bulge in M31 with $R \approx 2.4-2.5$ \citep{2014ApJ...785..136D}.
	The green dotted line is the average of the four extinction curves derived by \citet{2015ApJ...815...14C}.
	The $R_V$ values of MCs are given by \citet{2003ApJ...594..279G}.
	\label{fig:M31ext}}
\end{figure}

Each extinction curve for the selected tracers with \emph{pcFlag = `UVI'} is presented in Figure \ref{fig:M31ext} with a gray line, compared with those of the MW (black dashed line), LMC (blue dashed-and-dotted line), SMC (orange dashed-and-dotted line for the bar region and yellow dashed line for the wing region) and previous works on M31.
As illustrated in Figure \ref{fig:M31ext}, the extinction curves in M31 cover a wide range of shapes, from as flat as curves with large $R_V$ values, to steep ones with obvious 2175 $\,{\rm \AA}$ bumps, indicating the complex interstellar environment and inhomogeneous dust distribution along the M31 arms.
The red solid line shows the average extinction curve calculated by the median value of the dust size parameter $\alpha$, and is similar to those for the diffuse region in the MW and the average LMC, but with a slightly less steep far-UV rise.
The average extinction law of M31 derived in this work can be applied to the general extinction correction in M31.
The extinction curve for individual tracers can help with higher-precision extinction correction.
The details are interpreted in Section \ref{subsec:pre}.
It should be noted that there exists a second UV bump at $\sim$ 6.8 $\mu{\rm m}^{-1}$ in each derived extinction curve, which is not present in mathematical extinction curves.
This second UV bump is due to the sharp absorption edge of the ``astronomical silicate" dielectric function in \citet{1984ApJ...285...89D} based on laboratory measurements of crystalline olivine (Mg, Fe)$_{2}$SiO$_{4}$ in \citet{1973IAUS...52..297H}.
The dielectric function for ``astronomical silicate" in \citet{1984ApJ...285...89D} is typical and is the basis for modeling extinction laws with the silicate-graphite dust model \citep{2001ApJ...548..296W,2001ApJ...554..778L,2014P&SS..100...32W,2015ApJ...807L..26G,2020P&SS..18304627G}.

\begin{figure}
	\centering
	\includegraphics[scale=0.8]{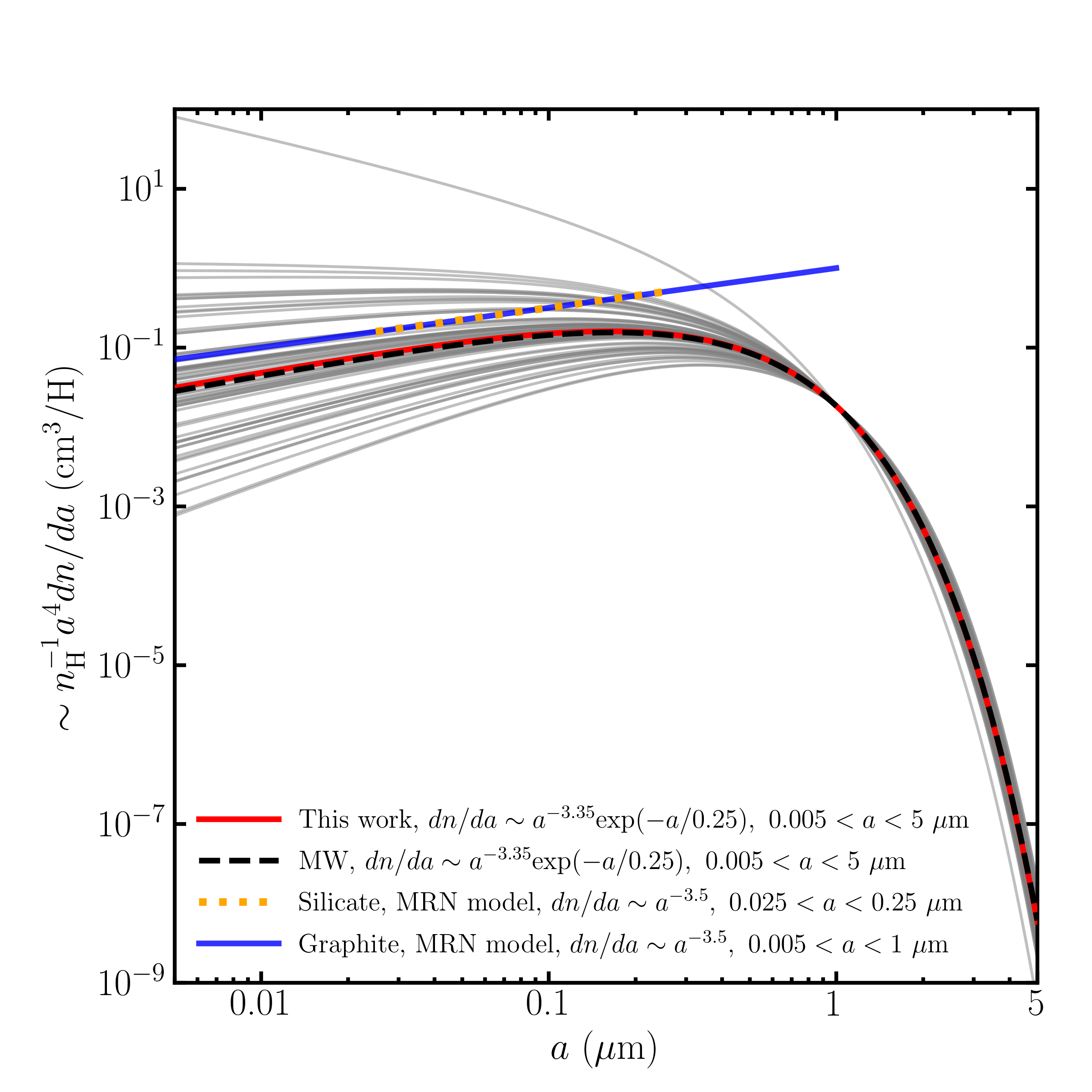}
	\caption{Comparison of the dust size distributions.
	The red solid line indicates the average dust size distribution for M31 derived in this work.
	The black dashed line shows the corresponding dust size distribution for the diffuse region in the MW as fitting model extinction curves to the CCM extinction law ($R_V = 3.1 $).
	The yellow dotted line and the blue solid line are the typical MRN dust size distributions for silicate and graphite, respectively \citep{1977ApJ...217..425M}.
	\label{fig:M31dis}}
\end{figure}

Since the single-parameter ($R_V$) extinction law cannot be universally applied in external galaxies, the parameter $\alpha$ in the dust size distribution is adopted in this work in order to describe the dust properties.
As presented in Table \ref{tab:re_dis}, the derived $\alpha$ that fits our model extinction curves to the average MW extinction curve ($\approx 3.33$, see the last line in Table \ref{tab:re_dis}) is close to the median $\alpha$ for the selected tracers with \emph{pcFlag = `UVI'} ($\approx 3.35$, see the first line in Table \ref{tab:re_dis}), indicating that the average dust size in M31 is similar to that of the diffuse region in the MW.
In addition, with $a_c$ in the KMH distribution fixed to 0.25 $\mu {\rm m}$, the derived $\alpha$ for the MW ($\approx 3.33$) is in the exponential range ($\approx 3.3-3.6$) of the MRN dust size distribution and is near the universal value ($\alpha = 3.5$), implying the reliability of our model extinction curves.
The dust size distribution for each of the selected tracers with \emph{pcFlag = `UVI'} is presented in Figure \ref{fig:M31dis} as a gray solid line with the average dust size distribution in M31 and the MW, as well as the typical MRN distributions for both silicate and graphite.
Despite the similarity between the average dust size distributions in M31 and the MW, various shapes of dust size distributions are shown in Figure \ref{fig:M31dis} that cannot be entirely interpreted by the single parameter $R_V$.

\movetabledown=1.5in
\begin{rotatetable}
	\begin{deluxetable}{ccccccccccccccccc}
		\tablecaption{An extracted list of the selected tracers in the sample with spectral type, fitting parameters [$\alpha,~{\rm log}(T_{\rm eff}),~{\rm log}(g),~A_V$], derived $E(B-V)$, $R_V$ and information about the observed photometry adopted in this work$^a$.
		\label{tab:re}}
		\tablehead{
		 \colhead{LGGS ID} & \colhead{LGGS SpT} & \colhead{$\alpha^{b,c}$} & \colhead{${\rm log}(T_{\rm eff})^c$} & \colhead{${\rm log}(g)^c$} & \colhead{$A_V$$^c$}  & \colhead{$E(B-V)$} & \colhead{$R_V$} &  \colhead{Total bands} & \colhead{pcFlag$^d$} & \colhead{...} \\	
		 \colhead{} & \colhead{} & \colhead{} & \colhead{} & \colhead{} & \colhead{(mag)}  & \colhead{(mag)} & \colhead{} &  \colhead{} & \colhead{} & \colhead{}
		}
		\startdata
        J004313.71+414245.3 & B2.5Ia & $2.76_{-0.04}^{+0.05}$ & $4.23_{-0.01}^{+0.01}$ & $2.55_{-0.07}^{+0.07}$ & $1.25_{-0.03}^{+0.03}$ & 0.21 & 5.92 & 13 & VI & ... \\
		J004341.45+410727.5 & B5I: & $3.68_{-0.15}^{+0.10}$ & $4.17_{-0.00}^{+0.01}$ & $2.5_{-0.08}^{+0.08}$ & $0.97_{-0.06}^{+0.08}$ & 0.35 & 2.8 & 13 & VI  & \\
		J004044.03+405238.2 & B8I & $3.40_{-0.07}^{+0.07}$ & $4.13_{-0.03}^{+0.02}$ & $2.26_{-0.09}^{+0.08}$ & $1.07_{-0.05}^{+0.08}$ & 0.32 & 3.39 & 15 & UVI  & \\
		J004017.61+405137.0 & B2.5I & $3.16_{-0.05}^{+0.07}$ & $4.28_{-0.01}^{+0.04}$ & $2.57_{-0.09}^{+0.15}$ & $1.47_{-0.05}^{+0.08}$ & 0.36 & 4.08 & 10 & V & \\
		J004350.18+412331.1 & B5I & $3.66_{-0.04}^{+0.07}$ & $4.22_{-0.05}^{+0.0}$ & $2.50_{-0.08}^{+0.08}$ & $0.83_{-0.16}^{+0.04}$ & 0.29 & 2.84 & 16 & UVI & \\
		J004246.86+413336.4 & O3-5If & $3.13_{-0.05}^{+0.04}$ & $4.60_{-0.04}^{+0.00}$ & $3.51_{-0.09}^{+0.08}$ & $2.96_{-0.04}^{+0.36}$ & 0.71 & 4.19 & 10 & UVI  \\
		J004130.12+405059.2 & O9.7Ia & $4.16_{-0.04}^{+0.04}$ & $4.51_{-0.01}^{+0.01}$ & $3.25_{-0.08}^{+0.09}$ & $0.75_{-0.04}^{+0.03}$ & 0.34 & 2.20 & 13 & VI  \\
		J004530.60+420655.4 & O8.5I & $3.02_{-0.08}^{+0.06}$ & $4.53_{-0.06}^{+0.02}$ & $3.50_{-0.08}^{+0.09}$ & $1.18_{-0.06}^{+0.08}$ & 0.26 & 4.61 & 17 & UVI \\
        \enddata
	\tablecomments{$^a$ This is an extracted table of results for some randomly selected O-type and B-type supergiants with different spectral types.
	The entire table with photometry used in each band is available in machine-readable form.
	The portion is shown here for guidance regarding its form and content.\\
	$^b$ The dust size distribution is expressed as:
	$dn/da \sim a^{-\alpha}{\rm exp}(-a/0.25),~0.005 < a < 5~\mu {\rm m}$.\\
	$^c$ The final results and the uncertainties for each parameter are derived by 50\%, 16\% and 84\% of the parameter spaces from the EMCEE results.\\
	$^d$ {This is the flag for the coverage of passbands adopted in calculation.
	U = UV bands (here refers to the passbands bluer than the $U$ band).
	V = Visual band (here refers to the passbands from the $U$ to $y$ bands).
	I = IR band (here refers to the passbands redder than $y$ band).}
	}		
	\end{deluxetable}
\end{rotatetable}

\begin{deluxetable*}{cccccccccc}
	\tablecaption{General results of $\alpha$, $A_V$, $E(B-V)$ and $R_V$ with the corresponding upper and lower limits for M31$^a$. \label{tab:re_dis}}
		\tablehead{	
		\colhead{} & \colhead{Total Sample} &  \colhead{$\alpha$} & \colhead{$A_V$} & \colhead{$E(B-V)$} & \colhead{$R_V$} \\
		\colhead{} & \colhead{} &  \colhead{} & \colhead{(mag)} & \colhead{(mag)} & \colhead{}
		}
	\startdata
	    pcFlag = `UVI'$^b$ & 70 & $3.35_{-0.70}^{+1.48}$ & $1.05_{-0.64}^{+1.92}$ & $0.26_{-0.09}^{+0.45}$ & $3.51_{-1.61}^{+3.14}$ \\
		pcFlag = `UV' & 6 & $3.06_{-0.42}^{+0.39}$ & $1.32_{-0.55}^{+1.14}$ & $0.33_{-0.16}^{+0.06}$ & $4.45_{-1.18}^{+2.28}$ \\
		pcFlag = `VI' & 29 & $3.55_{-0.91}^{+1.51}$ & $1.05_{-0.74}^{+1.89}$ & $0.32_{-0.16}^{+0.71}$ & $3.05_{-1.15}^{+3.68}$\\
		pcFlag = `V' & 35 & $3.32_{-0.67}^{+2.62}$ & $0.94_{-0.50}^{+1.20}$ & $0.26_{-0.09}^{+0.31}$ & $3.59_{-1.68}^{+3.06}$\\
		All tracers & 140 & $3.37_{-0.73}^{+2.58}$ & $1.05_{-0.74}^{+1.92}$ & $0.27_{-0.11}^{+0.76}$ & $3.47_{-1.58}^{+3.25}$ \\
		\hline
		Tracers of O-type & 9 & $3.03_{-0.37}^{+1.13}$ & $1.18_{-0.42}^{+1.79}$ & $0.26_{-0.09}^{+0.77}$ & $4.57_{-2.37}^{+2.01}$ \\
		Tracers of B-type & 131 & $3.39_{-0.76}^{+2.55}$ & $1.01_{-0.70}^{+1.45}$ & $0.27_{-0.11}^{+0.31}$ & $3.41_{-1.51}^{+3.32}$ \\
		\hline
		Tracers with near-IR data$^c$ & 99 & $3.40_{-0.76}^{+1.66}$ & $1.05_{-0.74}^{+1.91}$ & $0.28_{-0.12}^{+0.76}$ & $3.39_{-1.49}^{+3.34}$ \\
		 & & $3.31_{-0.69}^{+1.12}$$^{\ast}$ & $1.13_{-0.77}^{+2.74}$$^{\ast}$ & $0.28_{-0.11}^{+0.76}$$^{\ast}$ & $3.61_{-1.59}^{+3.27}$$^{\ast}$ \\
		Tracers with UV data$^c$ & 76 & $3.35_{-0.70}^{+1.49}$ & $1.05_{-0.64}^{+1.91}$ & $0.26_{-0.09}^{+0.45}$ & $3.51_{-1.61}^{+3.22}$ \\
		 & & $3.41_{-0.80}^{+1.42}$$^{\ast}$ & $1.12_{-0.76}^{+1.63}$$^{\ast}$ & $0.31_{-0.14}^{+0.59}$$^{\ast}$ & $3.37_{-1.47}^{+3.59}$$^{\ast}$\\
		\hline
		MW (CCM, $R_V = 3.1$)$^d$ & & 3.33 & &  
	\enddata
	\tablecomments{$^a$ The superscript and the subscript in the table are the derived upper limit value and lower value of all the selected tracers based on Table \ref{tab:re}, respectively.\\
	$^b$ The results of the sightlines with \emph{pcFlag = `UVI'} are adopted to derive the general extinction law in M31.\\
	$^c$ For tracers with near-IR (UV) data, the calculation is repeated without taking near-IR (UV) data into consideration, and the results are listed with $^{\ast}$ for comparison.\\
	$^d$ The Levenberg-Marquardt method is adopted for fitting the model extinction curves to the CCM extinction curve with $R_V = 3.1$.
	There is only one parameter ($\alpha$) in our model extinction curves, and a grid ranging from 0.50 to 7.00 with a step of 0.01 is taken.\\
	}
\end{deluxetable*}

\newpage

\subsection{Comparison with Other Works}\label{subsec:com}

\begin{figure}
	\includegraphics[scale=1.5]{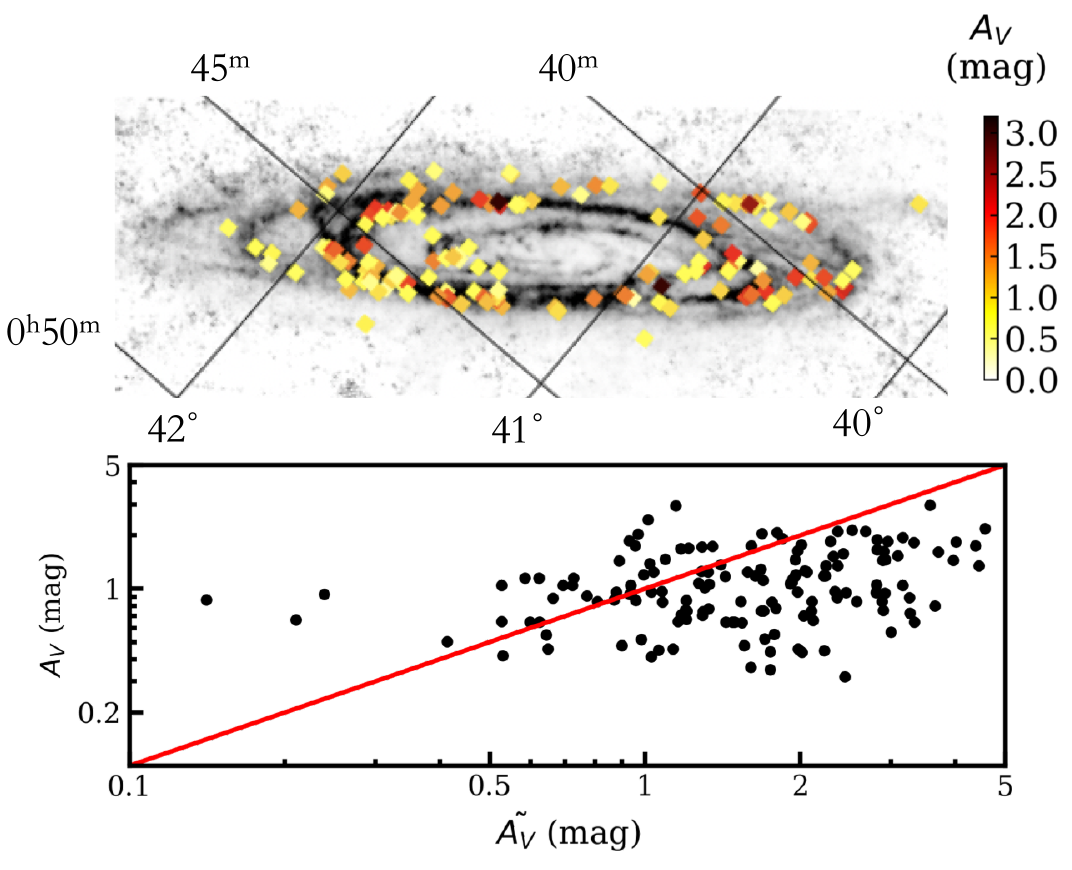}
	\caption{$A_V$ derived in this work against the dust mass surface density map in \citet{2014ApJ...780..172D} (the upper panel) and the derived $A_V$ versus the theoretical $\tilde{A_V}$ inferred from the dust emission in \citet{2014ApJ...780..172D} (the lower panel).
	The empirical relationship between $\tilde{A_V}$ and the dust mass surface density inferred from the dust emission is $\tilde{A_V} = 0.74(\frac{\sum_{M\rm d}}{10^5M\odot \rm{kpc^{-2}}})$ \citep{2007ApJ...657..810D}.
	 \label{fig:com_draine} }
\end{figure}

A comparison of the extinction curves in M31 derived in this work with other previous extinction works is also shown in Figure \ref{fig:M31ext}.
Each extinction curve for the selected tracers in the sample is presented as a gray solid line, while the red solid line is the average extinction law.
Some flat extinction curves of M31 with larger values of $R_V$ and weaker 2175 $\,{\rm \AA}$ bumps resemble the extinction law derived by \citet{1996ApJ...471..203B} shown in Figure \ref{fig:M31ext} with cyan triangles.
The purple squares in Figure \ref{fig:M31ext} are the average extinction law for the five dusty clumps located in the circumnuclear region with $R_V \approx 2.4 - 2.5$ derived by \citet{2014ApJ...785..136D}, which is steeper than that of the diffuse region in the MW.
The derived values of $A_V$ for each selected tracer against the dust mass surface density map derived by \citet{2014ApJ...780..172D} are plotted in Figure \ref{fig:com_draine}.
Although nearly all the selected tracers in this work are distributed along the arms of M31, as shown in Figure \ref{fig:com_draine}, some extinction curves are as steep as the average extinction law in \citet{2014ApJ...785..136D}, and a few are steeper, indicating the complex interstellar environment along the arms in M31.
In addition, the average extinction curve in this work is flatter than the extinction law in \citet{2014ApJ...785..136D}, from which it can be inferred that there are larger dust grains along the arms than those in the circumnuclear region.
The green dotted line in Figure \ref{fig:M31ext} is the average result of the four extinction curves derived in \citet{2015ApJ...815...14C} with $R_V \approx 2.5$, showing a similarity to the average extinction curve of M31 in the far-UV and visual bands, but with a slightly stronger 2175 $\,{\rm \AA}$ bump.
The wide range of extinction curves in this work corresponds to the different extinction curves ($R_V \approx 2 - 3.3$) towards different sightlines presented in \citet{2015ApJ...815...14C}, both illustrating the inhomogeneity of dust distribution in M31.

\begin{figure}
	\includegraphics[scale=0.625]{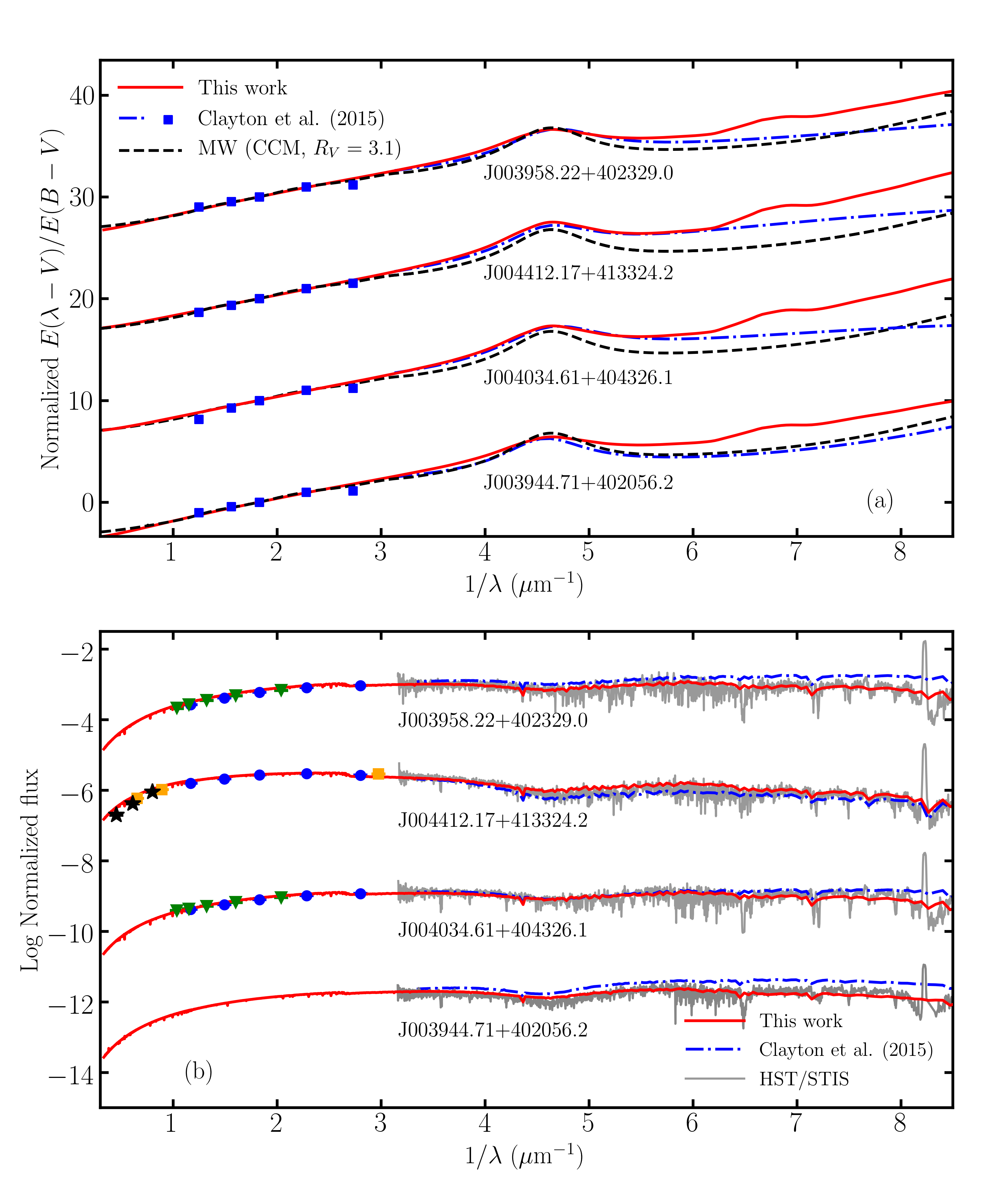}
	\caption{Panel (a): comparison of the extinction curves towards the individual sightlines derived in this work (red solid lines ) and those (blue dashed-and-dotted lines for UV bands and blues squares for $UBVRI$) in \citet{2015ApJ...815...14C}.
	Panel (b): comparison of the model SEDs for the four tracers derived in this work with the \emph{HST}/STIS spectra and the observed photometry (the symbol convention follows Figure \ref{fig:example}).
	The blue dashed-and-dotted lines indicate the model SEDs constructed with the derived parameters in \citet{2015ApJ...815...14C}.
	 \label{fig:com_Clayton} }
\end{figure}

The sample of the extinction tracers in this work contains two tracers of the four tracers in \citet{2015ApJ...815...14C}, i.e. J003958.22+402329.0 and J004412.17+413324.2.
Figure \ref{fig:com_Clayton} (a) compares the extinction curves towards the two tracers derived in this work and those derived in \citet{2015ApJ...815...14C}.
Comparison between the extinction curves towards the other two tracers derived with the method adopted in this work and derived in \citet{2015ApJ...815...14C} are also shown in Figure \ref{fig:com_Clayton} (a).
For the tracer J004034.61+404326.1, it is a B-type supergiant according to the LGGS catalog, but the result derived with the photometric data is considered to be unreliable based on the selection criterion that $R_V =1.5-7$ mentioned in Section \ref{subsec:selection}, which may due to lacking limitation of the UV data (see Section \ref{subsec:rv} for further discussions).
So we combine the Hubble Space Telescope/Space Telescope Imaging Spectrograph (\emph{HST}/STIS) spectra adotped in \citet{2015ApJ...815...14C} with the observed photometry to derive the extinction curve using the same method.
As to the fourth tracer J003944.71+402056.2, it may be a crowded source (Cwd = `C' in the LGGS catalog) and cannot be checked with the PHAT/F475W image, so we just adopt the \emph{HST}/STIS spectra to calculate the extinction curve.
In Figure \ref{fig:com_Clayton} (b), the model SEDs for the four tracers derived in this work are compared with the \emph{HST}/STIS spectra and the observed photometry.
The model SEDs with the stellar parameters and extinction parameters derived in \citet{2015ApJ...815...14C} are also constructed in Figure \ref{fig:com_Clayton} (b).
It should be noted that the values of surface gravity for the tracers J003944.71+402056.2 [log$(g) = $ 2.79] and J003958.22+402329.0 [log$(g) = $ 2.08] presented in \citet{2015ApJ...815...14C} are not in the range of the Tlusty grid.
As a result, we adopt the nearest log$(g)$ values [log$(g) = $ 3 for J003944.71+402056.2 and log$(g) = $ 2.5 for J003958.22+402329.0] to model the intrinsic spectra with the Tlusty stellar model atmosphere.
We can see the consistency in these extinction curves towards the individual sightlines in near-UV and optical bands from Figure \ref{fig:com_Clayton} (a), but deviation exists in far-UV bands.
It is shown in Figure \ref{fig:com_Clayton} (b) that the model SEDs calculated in this work fit the \emph{HST}/STIS spectra and the observed photometry well.

The results of the four individual sightlines derived in this work are presented in Table \ref{tab:com_clayton} with the comparison of those dervied in \citet{2015ApJ...815...14C}.
Because of the different extinction models, the values of $R_V$ derived in this work are larger than those in \citet{2015ApJ...815...14C}, indicating that $R_V$ cannot be adopted to completely describe the extinction curves in M31 (see Section \ref{subsec:rv} for further discussions).


\movetabledown=1.5in
\begin{rotatetable}
	\begin{deluxetable}{ccccccccccccccccc}
		\tablecaption{The results of the four individual sightlines derived in this work compared with those in \citet{2015ApJ...815...14C}$^a$.
		\label{tab:com_clayton}}
		\tablehead{
		 \colhead{LGGS ID} & \colhead{LGGS SpT} & \colhead{$\alpha$} & \colhead{${\rm log}(T_{\rm eff})$} & \colhead{${\rm log}(g)$} & \colhead{$A_V$$^b$}  & \colhead{$E(B-V)^c$} & \colhead{$R_V$$^d$} &  \colhead{Observed data}  \\	
		 \colhead{} & \colhead{} & \colhead{} & \colhead{} & \colhead{} & \colhead{(mag)}  & \colhead{(mag)} & \colhead{}
		}
		\startdata
		J003958.22+402329.0 & B0.7 Ia & 3.40 & 4.45 & 2.99 & 1.06 & 0.32 & 3.30 & Photometry\\
		  & & - & 4.35 & 2.08 & 0.60 & 0.24 & 2.50 & \emph{HST}/STIS \\		
		J004412.17+413324.2 & B2.5 Ia & 3.54 & 4.36 & 2.53 & 1.55 & 0.51 & 3.10 & Photometry\\
		 & & - & 4.24 & 2.32 & 0.78 & 0.39 & 2.00 & \emph{HST}/STIS\\		  	  
		J004034.61+404326.1 & B1 Ia & 3.50 & 4.34 & 2.74 & 1.10 & 0.34 & 3.20 & Photometry + \emph{HST}/STIS\\
		 & & - & 4.35 & 2.53 & 0.63 & 0.25 & 2.50 & \emph{HST}/STIS \\
        J003944.71+402056.2 & O9.7 Ib & 3.30 & 4.47 & 3.25 & 1.16 & 0.32 & 3.70 & \emph{HST}/STIS \\
		 & & - & 4.48 & 2.79 & 1.22 & 0.37 & 3.30 & \emph{HST}/STIS \\
        \enddata
	\tablecomments{$^a$ For a certain tracer, the first line shows the results derived in this work, while the second line is the results from \citet{2015ApJ...815...14C}.\\
	$^b$ $A_V$ in this work is the 50\% values of the posterior probability distribution functions generated from the EMCEE results. 
	While $A_V$ in \citet{2015ApJ...815...14C} is calculated by $A_V = R_V \times E(B-V)$.\\
	$^c$ $E(B-V)$ in this work is obtained based on the derived extinction curves as mentioned in Section \ref{subsec:eachre}.
	While $E(B-V)$ from \citet{2015ApJ...815...14C} is extracted from Table 5 of the paper.\\
	$^d$ $R_V$ in this work is calculated by $R_V = A_V/E(B-V)$, and that from \citet{2015ApJ...815...14C} refers to Section 5 in their paper.}
	\end{deluxetable}
\end{rotatetable}


\citet{2014ApJ...780..172D} mapped the dust mass surface density $\sum_{M\rm d}$ by modeling the observed far-IR and submillimeter emission.
The dust model adopted in \citet{2014ApJ...780..172D} is as described in \citet[hereafter DL07]{2007ApJ...657..810D}, in which the dust is assumed to be a mixture of carbonaceous grains and amorphous silicate grains 
with a size distribution proposed by \citet{2001ApJ...548..296W}.
According to DL07, the relationship between the extinction in the V band and the dust mass surface density $\sum_{M\rm d}$ is $\tilde{A_V} = 0.74(\frac{\sum_{M\rm d}}{10^5M\odot \rm{kpc^{-2}}})$ mag.
We then plot the $A_V$ derived in this work against the dust map in the upper panel of Figure \ref{fig:com_draine} and show the derived $A_V$ versus the $\tilde{A_V}$ calculated from the dust mass surface density in the lower panel of Figure \ref{fig:com_draine}.
From figure \ref{fig:com_draine}, we can determine that a number of $\tilde{A_V}$ values are larger than those derived in this work.

\citet{2015ApJ...814....3D} mapped the dust extinction in M31 by modeling the CMD of the red giant branch (RGB) stars using photometry from the PHAT survey and derived that the median value of $A_V$ is 1 mag, with only 10 \% pixels having $A_V > 1.8$ mag, 1 \% having $A_V > 2.8$ mag and very little surface area having $A_V > 3$ mag.
In this work, the median value of extinction is $A_V \approx 1.05$ mag, 18 of the 140 selected tracers ($\approx$ 12.9 \%) have $A_V > 1.8$ mag, 2 tracers ($\approx$ 1.4 \%) have $A_V > 2.8$ mag, and only no tracer has $A_V > 3$ mag.
\citet{2015ApJ...814....3D} argues that the true dust mass of M31 is lower and that dust grains are significantly more emissive than assumed in \citet{2014ApJ...780..172D}.
They also suggest that the DL07 model adopted in \citet{2014ApJ...780..172D} overpredicts the extinction in M31 by a factor of $\approx 2.5$.
Similarly, we reveal a fraction of $\approx 1.9$ offset between the $A_V$ directly derived in this work and those inferred from dust emission in \citet{2014ApJ...780..172D}.
The discrepancy may come from the different methods adopted and the distinct resolutions of dust.


\subsection{The Parameter $R_V$} \label{subsec:rv}

In this work, $R_V$ is obtained by $R_V = A_V/E(B-V)$.
It is universally known that $R_V$ may indicate the dust size along the sightline.
Theoretically, the steep extinction curve caused completely by Rayleigh scattering has $R_V \approx 0.72$, while $R_V$ could be infinite in an extremely dense region with very large, ``gray'' grains \citep{2003ARA&A..41..241D}, where the extinction does not vary much with the wavelength.
Observationally, as mentioned in Section \ref{sec:intro}, $R_V$ could be $2 - 6$ in the MW \citep{1990ARA&A..28...37M,1999PASP..111...63F,2017ApJ...848..106W,2019ApJ...877..116W}.

In this work there are 20 significantly reddened tracers in the sample with a good MCMC performance, yet the derived $R_V$ values are unusually high ($R_V > 7$).
Although the results lead to acceptable models that fit the observations well, the derived $R_V$ values are beyond the observational range of $R_V$ in the MW.
These results are consequently considered to be unreliable.

$R_V$ is the only parameter in the CCM model \citep{1989ApJ...345..245C}.
Extinction curves with a smaller $R_V$ usually have stronger 2175 $\,{\rm \AA}$ bumps and steeper far-UV rises.
However, $R_V$ cannot completely describe the extinction features on the extinction curve in this work, and it is better to use dust models with different dust size distributions to model the extinction curves.
It is generally accepted that the carrier of the 2175 $\,{\rm \AA}$ bump is a mixture of polycyclic aromatic hydrocarbons (PAHs) \citep{1992ApJ...393L..79J,2001ApJ...554..778L,2011ApJ...733...91X,2011ApJ...742....2S,2015ApJ...809..120M,2017ApJ...850..138M}.
The strength of the 2175 $\,{\rm \AA}$ bump depends on both the abundance and the size distribution of carbonaceous grains, rather than $R_V$.
For instance, \citet{2013ApJ...771...68P} found that no single $R_V$ parameter fits both the optical and UV extinction at high latitude in the MW.
In addition, although they have similar $R_V$ values, the extinction curve of the SMC bar ($R_V
\approx 2.74$) varies substantially from that of the SMC wing ($R_V \approx 2.05$) (see Figure \ref{fig:M31_intro} and Figure \ref{fig:M31ext}).
Therefore, $R_V$ cannot generally be applied to external galaxies such as SMC \citep{1982A&A...113L..15L,1984A&A...132..389P,2003ApJ...594..279G}, starburst galaxies \citep{1994ApJ...429..582C,2001PASP..113.1449C,1997AJ....113..162C,2000ApJ...533..682C}, type Ia supernova \citep{2015ApJ...807L..26G,2020P&SS..18304627G}, etc.

In addition to fixing the mass ratio of graphite to silicate to $f_{cs} = 0.3$, another value of $f_{cs} = 0.6$ has been adopted to perform the same calculation in this work, which means that all carbon is in the solid phase, but it does not substantially change the derived extinction results.
Therefore, in the case of M31, the variation in abundance probably has little effect on the extinction law of M31, and mathematical extinction models such as CCM \citep{1989ApJ...345..245C} or FM90 \citep{1990ApJS...72..163F} may be applicable in M31, similar to those in MW.
However, as mentioned in Section \ref{subsec:eachre} and other previous works, the extinction curves cover a wide range of shapes from the central region to the outskirt in M31.
As a result, instead of $R_V$, the parameter $\alpha$ in the dust size distribution function is preferred to describe the extinction law and the dust properties in M31.
Based on the detailed dust model, the dust extinction curves derived in this work can be more physical and cover wider wavelength ranges.


\subsection{Parameter Sensitivities} \label{subsec:sensi_test}

\begin{figure}[ht!]
	\includegraphics[scale=0.7]{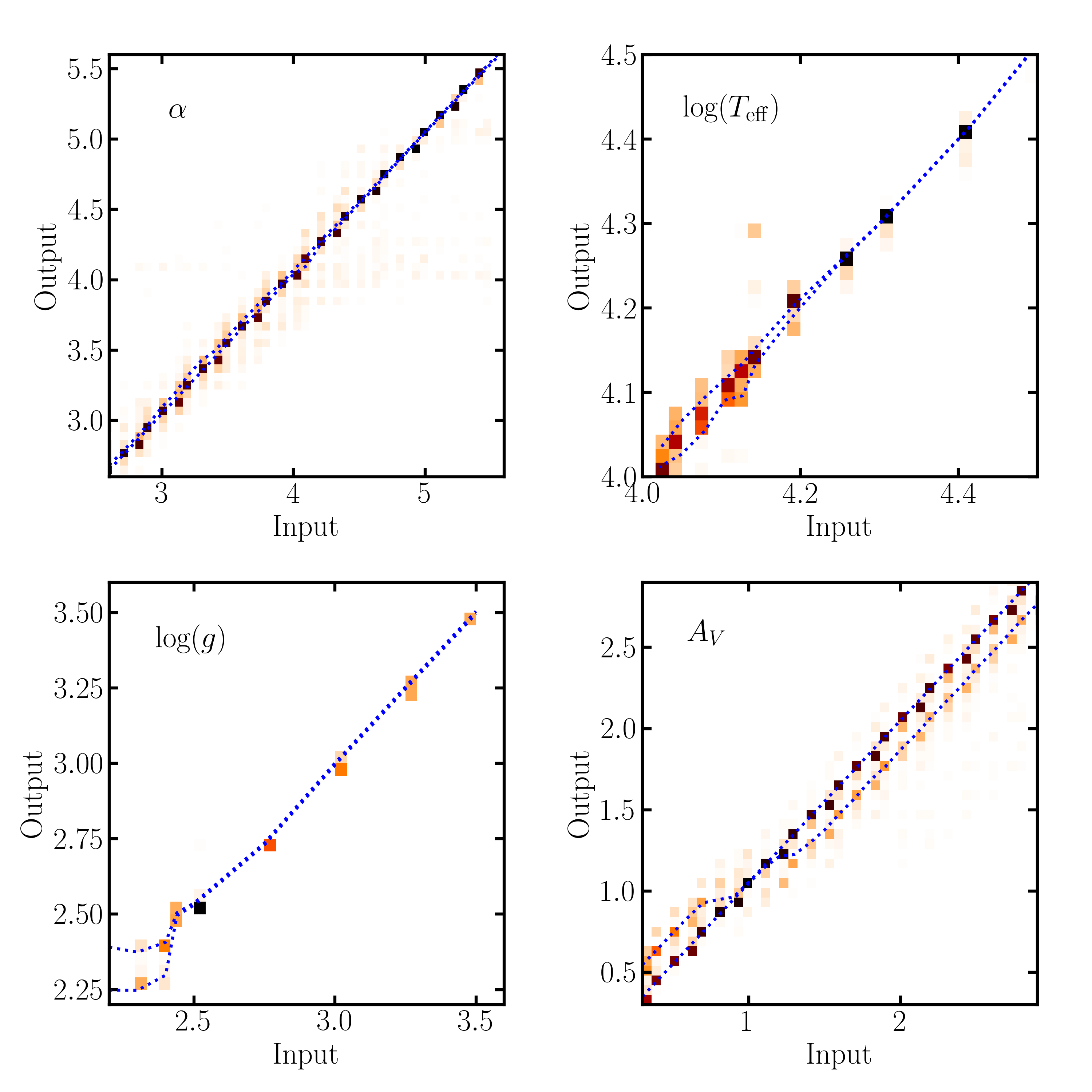}
	\caption{The results of sensitivity tests are shown with density plots for model SEDs consisting 9 photometric points ($UVW1$, $U$, $B$, $V$, $R$, $I$, $J$. $H$, $K$) extracted from the grid of model SEDs constructed in this work.
	The blue dotted lines give the 1 $\sigma$ (67 \%) regions.}
	 \label{fig:sensi_test} 
\end{figure}

The number of photometric points for more than 85 \% tracers in this work is at least 9.
In order to test how well the 4 model parameters can be recovered with such observed data, we test the parameter sensitivities.
For the sensitivity tests, the ``observed'' SEDs consisting 9 photometric points ($UVW1$, $U$, $B$, $V$, $R$, $I$, $J$, $H$, $K$, a common combination of observed photometry for the individual tracers in this work) with 33 values of $\alpha$ ($2.5 < \alpha < 5.9$ with 0.1 dex steps) and 27 values of $A_V$ ($0.2 < A_V < 3.0$ with 0.1 mag steps) for O-type and B-type supergiants with 12 spectral types (O8-9, B0-9) are picked from the grid of model SEDs constructed in Section \ref{subsec:fullsed}.   
Figure \ref{fig:sensi_test} shows the resulting recovery of parameters.

Figure \ref{fig:sensi_test} presents good recovery of the parameters.
The results of the sensitivity tests indicate that the results derived in this work are not strongly affected by the degeneracies between log($T_{\rm eff}$) and $A_V$ because of the wide wavelength coverage and the known distance to the tracers \citep{2016ApJ...826..104G}.
In addition, the Gaussian prior imposed on the log($T_{\rm eff}$) places restrictions on the range of derived log($T_{\rm eff}$), so that the impact of the degeneracies bewteen log($T_{\rm eff}$) and $A_V$ can be reduced.

\subsection{Influence of IR and UV Photometry} \label{subsec:IRUV}

Photometry in more bands will better constrain the observed SED and bring more reliable results.
Because of the observation limit, a number of tracers lack photometry in the $UVW2, UVM2, UVW1$ and PHAT bands.
Meanwhile, UKIRT data are also not applied to all tracers in the extinction sample, as mentioned in Section \ref{subsec:UKIRT}.
It is thus necessary to determine whether the lack of photometry in the UV bands and in the near-IR bands affects the derived extinction law.

As illustrated in Section \ref{subsec:selection}, unreliable results of individual tracers have been eliminated in this work to reduce the influence on the final results.
Some tracers with a larger non-physically-derived $R_V$ lack UV photometry, indicating that lacking UV data may lead to poor results because the UV data can effectively constrain the shape of the extinction curve near the 2175 $\,{\rm \AA}$ bump.
For early-type supergiants with high effective temperatures, if the observed data in UV or shorter bands are available to restrict the shape of the entire SED, the results will be more reliable.
In addition, all the tracers with acceptance fractions beyond $0.2 - 0.5$ lack UV and near-IR photometry or have a small number of fitting bands (approximately 10 or less), implying that insufficient data may bring about a negative MCMC performance.

\begin{figure}[ht!]
	\includegraphics[scale=0.4]{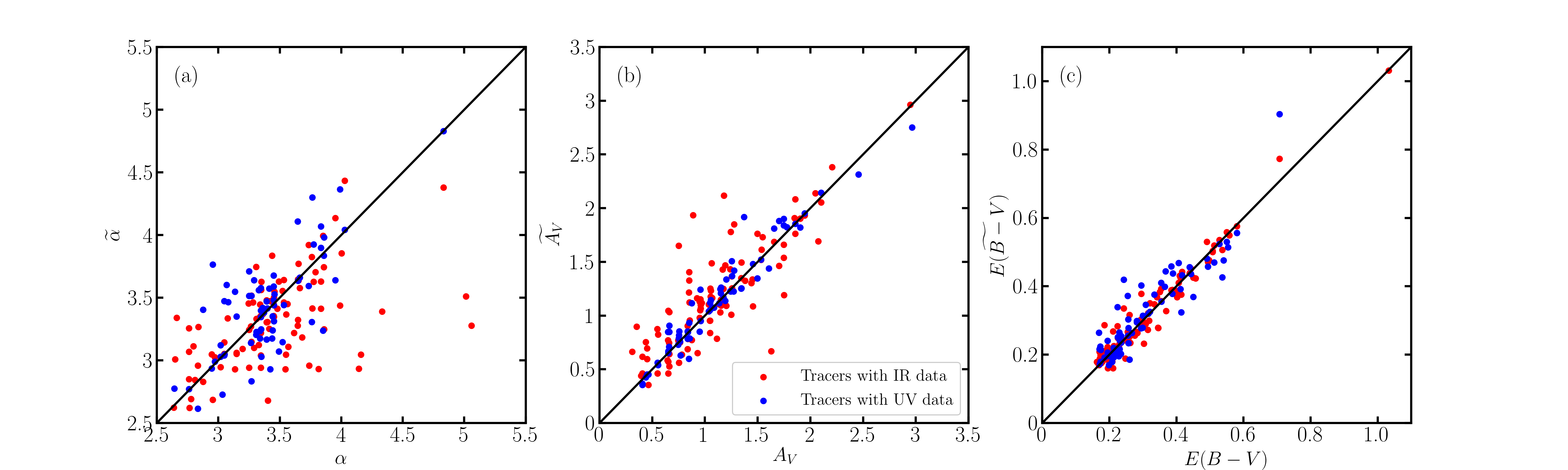}
	\caption{Influence of lacking UV or near-IR data on the derived $\alpha,~A_V$ and $E(B-V)$.
	The red and blue dots indicate the tracers with near-IR data and UV data, respectively.
	The values of x-axis for a certain red (blue) dot are the derived parameters [$\alpha,~A_V,~E(B-V)$] with all available photometric data considered, while the corresponding values of y-axis are the dervied parameters [$\widetilde{\alpha},~\widetilde{A_V},~\widetilde{E(B-V)}$] when repeat the calculation process but without taking the available near-IR (UV) data into consideration.
	The deviation between the red (blue) dots and $y=x$ presents the influence of the lack of near-IR (UV) data.
	 \label{fig:av_dis} }
\end{figure}

For the remained tracers with near-IR (UV) photometry, the results are summarized in the 8th (10th) line of Table \ref{tab:re_dis}, while the 9th (11th) line shows the results for the same tracers, but ignores the near-IR (UV) data.
Figure \ref{fig:av_dis} compares the values of $\alpha$, $A_V$ and $E(B-V)$ extracted from Table \ref{tab:re} with those derived without taking near-IR (UV) data into consideration.
From Table \ref{tab:re_dis} and Figure \ref{fig:av_dis}, we can see the consistency of $A_V$ and $E(B-V)$, implying that a lack of UV or near-IR data has little impact on $A_V$ and $E(B-V)$ in this work.
This is probably because all the tracers have photometry in the $U$ band, providing the preliminary constraint in the UV bands.
However, the dust size parameter $\alpha$ of the individual tracers is influenced by the coverage of the adopted passbands, indicating that the UV and IR data play an integral role in constraining the dust model.

Lacking UV or near-IR data may lead to poor constraint in the fits.
In order to illustrate the reliability of the derived re sults for individual sightlines, as mentioned in Section \ref{sec:data}, we introduce \emph{pcFlag} in Table \ref{tab:re} to show the coverage of passbands adopted in the calculation.
Results of sightlines with \emph{pcFlag = `UVI'} are the most reliable, while those with \emph{pcFlag = `V'} are the least reliable.
It is anticipated that if observed data in various bands are adequate, the results could be more comprehensive.
In the near future the coming 2 m-aperture Survey Space Telescope (also known as the China Space Station Telescope, CSST) will image approximately 17500 square degrees of the sky in the $NUV$, $u$, $g$, $r$, $i$, $z$ and $y$ bands \citep{2021CSB...111.111C} and will provide us with abundant data to explore the dust extinction law in M31 and other nearby star-resolved galaxies.


\subsection{Prediction of Multi-band Extinction} \label{subsec:pre}

The dust extinction law derived in this work can provide multiband extinction values and help with the extinction correction in M31.
Based on the average extinction curve of M31 in this work, extinction values for multiple bands from UV to near-IR are predicted, which are shown in Table \ref{tab:pre}.
High-precision extinction correction should refer to the extinction curve of individual tracers derived in this work, which is partially presented in Table \ref{tab:pre_indi}.
This is an extracted list for a few tracers as examples.
The entire table for all the selected tracers with multiband extinction values is available in machine-readable form.
The first four columns are the ID, spectral type, right ascension, and declination obtained from the LGGS catalog \citep{2016AJ....152...62M}, respectively.
The fifth and sixth columns present the derived $\alpha$ and $E(B-V)$ extracted from Table \ref{tab:re}.
The column named \emph{pcFlag} presents the passband coverage, as shown in Table \ref{tab:re}.
The following columns show the extinction values in multiple bands for each tracer.

However, the size of the extinction sample adopted in this work is not sufficiently adequate to cover the entire region of M31.
Thus, it can only provide us with a low-resolution extinction map (see the upper panel of Figure \ref{fig:com_draine}) that helps us with a rough extinction correction for certain regions in M31.
The coming CSST \citep{2021CSB...111.111C} mentioned in Section \ref{subsec:IRUV} is a major science project of the China Manned Space Program and will perform high-resolution large-area multiband imaging and slitless spectroscopy survey covering the wavelength range of 255 - 1000 nm.
The dust extinction tracers will then be enlarged
, which is advantageous for the further exploration of the dust properties and extinction law of M31 and other nearby star-resolved galaxies and the development of higher-precision extinction corrections in the near future.

\startlongtable
\begin{deluxetable}{cccccccccc}
	\tablecaption{General prediction of extinction in M31 for multibands from UV to IR. \label{tab:pre}}
		\tablehead{	
		\colhead{Band} & \colhead{$\lambda_{\rm eff}$$^a$} & \colhead{$A_{\lambda}/A_V$} & \colhead{$A_{\lambda}$$^b$ } \\
		\colhead{} & \colhead{$(\mu {\rm m})$} & & \colhead{(mag)} 
		}
	\startdata	                
	$UVW2$/UVOT & 0.209 & $2.85$ & $2.99$ \\
	$UVM2$/UVOT & 0.225 & $2.78$ & $2.92$ \\
	$UVW1$/UVOT & 0.268 & $2.11$ & $2.21$ \\
	$F275W$/PHAT & 0.272 & $2.07$ & $2.16$ \\
	$NUV$/CSST & 0.29 & $1.92$ & $2.01$ \\
	$F336W$/PHAT & 0.336 & $1.65$ & $1.73$ \\
	$U$ & 0.357 & $1.55$ & $1.63$ \\
	$B$ & 0.443 & $1.26$ & $1.31$ \\
	$F475W$/PHAT & 0.473 & $1.17$ & $1.23$ \\
	$g$/CSST & 0.475 & $1.17$ & $1.22$ \\
	$V$ & 0.554 & $1.00$ & $1.05$ \\
	$r$/CSST & 0.612 & $0.89$ & $0.93$ \\
	$R$ & 0.67 & $0.8$ & $0.84$ \\
	$i$/CSST & 0.758 & $0.7$ & $0.73$ \\
	$F814W$/PHAT & 0.798 & $0.65$ & $0.68$ \\
	$I$ & 0.857 & $0.6$ & $0.63$ \\
	$z$/CSST & 0.911 & $0.55$ & $0.58$ \\
	$y$/CSST & 0.989 & $0.5$ & $0.52$ \\
	$F110W$/PHAT & 1.12 & $0.42$ & $0.44$ \\
	$J$/2MASS & 1.235 & $0.36$ & $0.38$ \\
	$F160W$/PHAT & 1.528 & $0.26$ & $0.28$ \\
	$H$/2MASS & 1.662 & $0.23$ & $0.24$ \\
	$K$/2MASS & 2.159 & $0.15$ & $0.16$ \\
	$W1$/WISE & 3.353 & $0.07$ & $0.07$ \\
	$[3.6]$/IRAC & 3.508 & $0.06$ & $0.06$ \\
	$[4.5]$/IRAC & 4.437 & $0.04$ & $0.04$ \\
	$W2$/WISE & 4.603 & $0.04$ & $0.04$
\enddata
	\tablecomments{$^a$ Effective wavelengths of multiple bands (except CSST bands) used in this work refer to the SVO Filter Profile Service (http://svo2.cab.inta-csic.es/theory/fps/, \citealt{2012ivoa.rept.1015R}).
	Effective wavelengths of CSST bands are calculated by $\lambda_{\rm eff} = \frac{\int \lambda^2T(\lambda)Vg(\lambda)d\lambda}{\int \lambda T(\lambda) Vg(\lambda)d \lambda}$, where $T(\lambda)$ is the filter transmission function and $Vg(\lambda)$ is the Vega spectrum.\\
	$^b$ $A_{\lambda}$ is the average extinction derived in this work for M31 based on the median value of $A_V$.}
	\end{deluxetable}
\newpage

\movetabledown=1.5in
\begin{rotatetable}
	\begin{deluxetable}{ccccccccccccccccc}
		\tablecaption{An extracted list of multiband extinction values for individual tracer$^a$. \label{tab:pre_indi}}
		\tablehead{	
		\colhead{LGGS ID} & \colhead{LGGS SpT} & \colhead{RA}  & \colhead{Dec} & \colhead{$\alpha$} & \colhead{$E(B-V)$} & \colhead{pcFlag$^b$} & \colhead{$A_V$} & \colhead{$A_{UVW2}$}  & \colhead{$A_{UVM2}$} & \colhead{$A_{UVW1}$} & \colhead{$A_{F275W}$} &  \colhead{...}\\
		\colhead{} & \colhead{} & \colhead{(h:m:s)} & \colhead{(d:m:s)} & \colhead{} &\colhead{(mag)} & \colhead{} & \colhead{(mag)} & \colhead{(mag)} & \colhead{(mag)} & \colhead{(mag)} & \colhead{(mag)}
		} 
\startdata
J004313.71+414245.3 & B2.5Ia & 00 43 13.70 & +41 42 45.1 & 2.76 & 0.21 & VI & 1.25 & 2.22 & 2.21 & 1.93 & 1.91 & ... \\
J004341.45+410727.5 & B5I: & 00 43 41.44 & +41 07 27.4 & 3.68 & 0.35 & VI & 0.97 & 3.78 & 3.62 & 2.47 & 2.40 &  ... \\
J004044.03+405238.2 & B8I & 00 40 44.02 & +40 52 38.1 & 3.40 & 0.32 & UVI & 1.07 & 3.20 & 3.11 & 2.32 & 2.27 & ... \\
J004017.61+405137.0 & B2.5I & 00 40 17.60 & +40 51 36.9 & 3.16 & 0.36 & V & 1.47 & 3.55 & 3.49 & 2.78 & 2.74 & ...  \\
J004350.18+412331.1 & B5I & 00 43 50.17 & +41 23 31.0 & 3.66 & 0.29 & UVI & 0.83 & 3.19 & 3.06 & 2.10 & 2.05 & ... \\
J004246.86+413336.4 & O3-5If & 00 42 46.85 & +41 33 36.3 & 3.14 & 0.71 & UVI & 2.96 & 6.98 & 6.88 & 5.53 & 5.44 & ... \\
J004130.12+405059.2 & O9.7Ia & 00 41 30.11 & +40 50 59.1 & 4.16 & 0.34 & VI & 0.75 & 4.66 & 4.33 & 2.48 & 2.39 &   ... \\
J004530.60+420655.4 & O8.5I & 00 45 30.59 & +42 06 55.2 & 3.02 & 0.26 & UVI & 1.18 & 2.53 & 2.50 & 2.07 & 2.04 &  ... \\
\enddata				
\tablecomments{$^a$ This is an extracted table for the same tracers as listed in Table \ref{tab:re} with a part of the multiband extinction values.
The entire table is available in machine-readable form.\\
$^b$ Flag for the coverage of passbands adopted in the calculation (the same as Table \ref{tab:re})}.
\end{deluxetable}
\end{rotatetable}


\newpage
\section{Conclusion} \label{sec:conclusion}

Based on the LGGS catalog, a sample of bright O-type and B-type supergiants is chosen as extinction tracers in M31.
These tracers are distributed along arms in M31.
The extinction sample size is larger than before; thus, new extinction curves towards more sightlines in M31 are derived.
The main results of this work are as follows:

1. The extinction curves of M31 derived in this work cover a wide range of shapes, from curves with an obvious 2175 $\,{\rm \AA}$ bump (like the MW extinction curves with $R_V \approx 2$) to relatively flat curves with $R_V \approx 6$, implying the complexity of the interstellar environment and the inhomogeneous distribution of interstellar dust in M31.
Some derived extinction curves look like those in former studies on M31 \citep{1996ApJ...471..203B,2014ApJ...785..136D,2015ApJ...815...14C}.
The derived parameter $\alpha$ in the dust size distribution ranges from $\approx 2.6-5.9$.

2. The average extinction curve of M31 ($R_V \approx 3.51$) shows a similarity to the CCM extinction curve with $R_V = 3.1$, but with a slightly less steep rise in far-UV bands.
The average dust size distribution in M31 is $dn/da \sim a^{-3.35}{\rm exp}(-a/0.25)$, similar to that of MW, indicating that the general interstellar environment in M31 may resemble the diffuse region in MW.

3. The derived $A_V$ in M31 is up to 3 mag with a median value of $\approx 1$ mag.
Most values ($\approx 90 \%$) of the derived extinction in the $V$ band are under 1.8 mag, which is in agreement with \citet{2015ApJ...814....3D}, but lower than the $\tilde{A_V}$ inferred from dust emission in \citet{2014ApJ...780..172D}.




\acknowledgments
We are grateful to Prof. Geoffrey Clayton for the very helpful comments and suggestions.
We greatly thank Profs. Biwei Jiang, Haibo Yuan, Wenyuan Cui, Drs. Shu Wang, Zhicun Liu, and Mr. Jun Li, Weijia Gao and Ruining Zhao for the very helpful discussions.
This work is supported by the National Natural Science Foundation of China through projects NSFC 12133002 and U2031209 and the CSST Milky Way and Nearby Galaxies Survey on Dust and Extinction Project CMS-CSST-2021-A09.
This work has made use of data from the LGGS, UKIRT, PS1 Survey, PHAT Survey, Swift/UVOT and XMM-SUSS.

%





\bibliography{paper}{}

\begin{thebibliography}{}
\expandafter\ifx\csname natexlab\endcsname\relax\def\natexlab#1{#1}\fi
\providecommand{\url}[1]{\href{#1}{#1}}
\providecommand{\dodoi}[1]{doi:~\href{http://doi.org/#1}{\nolinkurl{#1}}}
\providecommand{\doeprint}[1]{\href{http://ascl.net/#1}{\nolinkurl{http://ascl.net/#1}}}
\providecommand{\doarXiv}[1]{\href{https://arxiv.org/abs/#1}{\nolinkurl{https://arxiv.org/abs/#1}}}

\bibitem[{{Asplund} {et~al.}(2009){Asplund}, {Grevesse}, {Sauval}, \&
  {Scott}}]{2009ARA&A..47..481A}
{Asplund}, M., {Grevesse}, N., {Sauval}, A.~J., \& {Scott}, P. 2009, \araa, 47,
  481, \dodoi{10.1146/annurev.astro.46.060407.145222}

\bibitem[{{Bianchi} {et~al.}(1996){Bianchi}, {Clayton}, {Bohlin}, {Hutchings},
  \& {Massey}}]{1996ApJ...471..203B}
{Bianchi}, L., {Clayton}, G.~C., {Bohlin}, R.~C., {Hutchings}, J.~B., \&
  {Massey}, P. 1996, \apj, 471, 203, \dodoi{10.1086/177963}

\bibitem[{{Bianchi} {et~al.}(2012){Bianchi}, {Efremova}, {Hodge}, \&
  {Kang}}]{2012AJ....144..142B}
{Bianchi}, L., {Efremova}, B., {Hodge}, P., \& {Kang}, Y. 2012, \aj, 144, 142,
  \dodoi{10.1088/0004-6256/144/5/142}

\bibitem[{{Bless} \& {Savage}(1970)}]{1970IAUS...36...28B}
{Bless}, R.~C., \& {Savage}, B.~D. 1970, in IAU Symposium, Vol.~36, Ultraviolet
  Stellar Spectra and Related Ground-Based Observations, ed. R.~{Muller},
  L.~{Houziaux}, \& H.~E. {Butler}, 28

\bibitem[{{Boquien} {et~al.}(2010){Boquien}, {Duc}, {Galliano}, {Braine},
  {Lisenfeld}, {Charmandaris}, \& {Appleton}}]{2010AJ....140.2124B}
{Boquien}, M., {Duc}, P.~A., {Galliano}, F., {et~al.} 2010, \aj, 140, 2124,
  \dodoi{10.1088/0004-6256/140/6/2124}

\bibitem[{{Calzetti}(1997)}]{1997AJ....113..162C}
{Calzetti}, D. 1997, \aj, 113, 162, \dodoi{10.1086/118242}

\bibitem[{{Calzetti}(2001)}]{2001PASP..113.1449C}
---. 2001, \pasp, 113, 1449, \dodoi{10.1086/324269}

\bibitem[{{Calzetti} {et~al.}(2000){Calzetti}, {Armus}, {Bohlin}, {Kinney},
  {Koornneef}, \& {Storchi-Bergmann}}]{2000ApJ...533..682C}
{Calzetti}, D., {Armus}, L., {Bohlin}, R.~C., {et~al.} 2000, \apj, 533, 682,
  \dodoi{10.1086/308692}

\bibitem[{{Calzetti} {et~al.}(1994){Calzetti}, {Kinney}, \&
  {Storchi-Bergmann}}]{1994ApJ...429..582C}
{Calzetti}, D., {Kinney}, A.~L., \& {Storchi-Bergmann}, T. 1994, \apj, 429,
  582, \dodoi{10.1086/174346}

\bibitem[{{Cardelli} {et~al.}(1989){Cardelli}, {Clayton}, \&
  {Mathis}}]{1989ApJ...345..245C}
{Cardelli}, J.~A., {Clayton}, G.~C., \& {Mathis}, J.~S. 1989, \apj, 345, 245,
  \dodoi{10.1086/167900}

\bibitem[{{Castelli} \& {Kurucz}(2003)}]{2003IAUS..210P.A20C}
{Castelli}, F., \& {Kurucz}, R.~L. 2003, in Modelling of Stellar Atmospheres,
  ed. N.~{Piskunov}, W.~W. {Weiss}, \& D.~F. {Gray}, Vol. 210, A20.
\newblock \doarXiv{astro-ph/0405087}

\bibitem[{{Chaldu} {et~al.}(1973){Chaldu}, {Honeycutt}, \&
  {Penston}}]{1973PASP...85...87C}
{Chaldu}, R., {Honeycutt}, R.~K., \& {Penston}, M.~V. 1973, \pasp, 85, 87,
  \dodoi{10.1086/129408}

\bibitem[{{Chambers} {et~al.}(2016){Chambers}, {Magnier}, {Metcalfe},
  {Flewelling}, {Huber}, {Waters}, {Denneau}, {Draper}, {Farrow}, {Finkbeiner},
  {Holmberg}, {Koppenhoefer}, {Price}, {Rest}, {Saglia}, {Schlafly}, {Smartt},
  {Sweeney}, {Wainscoat}, {Burgett}, {Chastel}, {Grav}, {Heasley}, {Hodapp},
  {Jedicke}, {Kaiser}, {Kudritzki}, {Luppino}, {Lupton}, {Monet}, {Morgan},
  {Onaka}, {Shiao}, {Stubbs}, {Tonry}, {White}, {Ba{\~n}ados}, {Bell},
  {Bender}, {Bernard}, {Boegner}, {Boffi}, {Botticella}, {Calamida},
  {Casertano}, {Chen}, {Chen}, {Cole}, {Deacon}, {Frenk}, {Fitzsimmons},
  {Gezari}, {Gibbs}, {Goessl}, {Goggia}, {Gourgue}, {Goldman}, {Grant},
  {Grebel}, {Hambly}, {Hasinger}, {Heavens}, {Heckman}, {Henderson}, {Henning},
  {Holman}, {Hopp}, {Ip}, {Isani}, {Jackson}, {Keyes}, {Koekemoer}, {Kotak},
  {Le}, {Liska}, {Long}, {Lucey}, {Liu}, {Martin}, {Masci}, {McLean}, {Mindel},
  {Misra}, {Morganson}, {Murphy}, {Obaika}, {Narayan}, {Nieto-Santisteban},
  {Norberg}, {Peacock}, {Pier}, {Postman}, {Primak}, {Rae}, {Rai}, {Riess},
  {Riffeser}, {Rix}, {R{\"o}ser}, {Russel}, {Rutz}, {Schilbach}, {Schultz},
  {Scolnic}, {Strolger}, {Szalay}, {Seitz}, {Small}, {Smith}, {Soderblom},
  {Taylor}, {Thomson}, {Taylor}, {Thakar}, {Thiel}, {Thilker}, {Unger},
  {Urata}, {Valenti}, {Wagner}, {Walder}, {Walter}, {Watters}, {Werner},
  {Wood-Vasey}, \& {Wyse}}]{2016arXiv161205560C}
{Chambers}, K.~C., {Magnier}, E.~A., {Metcalfe}, N., {et~al.} 2016, arXiv
  e-prints, arXiv:1612.05560.
\newblock \doarXiv{1612.05560}

\bibitem[{{Chentsov} {et~al.}(2013){Chentsov}, {Klochkova}, {Panchuk},
  {Yushkin}, \& {Nasonov}}]{2013ARep...57..527C}
{Chentsov}, E.~L., {Klochkova}, V.~G., {Panchuk}, V.~E., {Yushkin}, M.~V., \&
  {Nasonov}, D.~S. 2013, Astronomy Reports, 57, 527,
  \dodoi{10.1134/S1063772913070019}

\bibitem[{{Cioni} {et~al.}(2008){Cioni}, {Irwin}, {Ferguson}, {McConnachie},
  {Conn}, {Huxor}, {Ibata}, {Lewis}, \& {Tanvir}}]{2008A&A...487..131C}
{Cioni}, M. R.~L., {Irwin}, M., {Ferguson}, A.~M.~N., {et~al.} 2008, \aap, 487,
  131, \dodoi{10.1051/0004-6361:200809366}

\bibitem[{{Clark} {et~al.}(2012){Clark}, {Najarro}, {Negueruela}, {Ritchie},
  {Urbaneja}, \& {Howarth}}]{2012A&A...541A.145C}
{Clark}, J.~S., {Najarro}, F., {Negueruela}, I., {et~al.} 2012, \aap, 541,
  A145, \dodoi{10.1051/0004-6361/201117472}

\bibitem[{{Clayton} {et~al.}(2015){Clayton}, {Gordon}, {Bianchi}, {Massa},
  {Fitzpatrick}, {Bohlin}, \& {Wolff}}]{2015ApJ...815...14C}
{Clayton}, G.~C., {Gordon}, K.~D., {Bianchi}, L.~C., {et~al.} 2015, \apj, 815,
  14, \dodoi{10.1088/0004-637X/815/1/14}

\bibitem[{{Clayton} \& {Martin}(1985)}]{1985ApJ...288..558C}
{Clayton}, G.~C., \& {Martin}, P.~G. 1985, \apj, 288, 558,
  \dodoi{10.1086/162821}

\bibitem[{{Conti} {et~al.}(2008){Conti}, {Crowther}, \&
  {Leitherer}}]{2008flhs.book.....C}
{Conti}, P.~S., {Crowther}, P.~A., \& {Leitherer}, C. 2008, {From Luminous Hot
  Stars to Starburst Galaxies}

\bibitem[{{Cox}(2000)}]{2000asqu.book.....C}
{Cox}, A.~N. 2000, {Allen's astrophysical quantities}

\bibitem[{{Cox}(1946)}]{1946AmJPh..14....1C}
{Cox}, R.~T. 1946, American Journal of Physics, 14, 1,
  \dodoi{10.1119/1.1990764}

\bibitem[{{Dalcanton} {et~al.}(2015){Dalcanton}, {Fouesneau}, {Hogg}, {Lang},
  {Leroy}, {Gordon}, {Sandstrom}, {Weisz}, {Williams}, {Bell}, {Dong},
  {Gilbert}, {Gouliermis}, {Guhathakurta}, {Lauer}, {Schruba}, {Seth}, \&
  {Skillman}}]{2015ApJ...814....3D}
{Dalcanton}, J.~J., {Fouesneau}, M., {Hogg}, D.~W., {et~al.} 2015, \apj, 814,
  3, \dodoi{10.1088/0004-637X/814/1/3}

\bibitem[{{Deng} {et~al.}(2020){Deng}, {Sun}, {Jian}, {Jiang}, \&
  {Yuan}}]{2020AJ....159..208D}
{Deng}, D., {Sun}, Y., {Jian}, M., {Jiang}, B., \& {Yuan}, H. 2020, \aj, 159,
  208, \dodoi{10.3847/1538-3881/ab8004}

\bibitem[{{Dong} {et~al.}(2016){Dong}, {Li}, {Wang}, {Lauer}, {Olsen}, {Saha},
  {Dalcanton}, \& {Groves}}]{2016MNRAS.459.2262D}
{Dong}, H., {Li}, Z., {Wang}, Q.~D., {et~al.} 2016, \mnras, 459, 2262,
  \dodoi{10.1093/mnras/stw778}

\bibitem[{{Dong} {et~al.}(2014){Dong}, {Li}, {Wang}, {Lauer}, {Olsen}, {Saha},
  {Dalcanton}, {Gordon}, {Fouesneau}, {Bell}, \&
  {Bianchi}}]{2014ApJ...785..136D}
---. 2014, \apj, 785, 136, \dodoi{10.1088/0004-637X/785/2/136}

\bibitem[{{Draine}(2003)}]{2003ARA&A..41..241D}
{Draine}, B.~T. 2003, \araa, 41, 241,
  \dodoi{10.1146/annurev.astro.41.011802.094840}

\bibitem[{{Draine}(2011)}]{2011piim.book.....D}
---. 2011, {Physics of the Interstellar and Intergalactic Medium}

\bibitem[{{Draine} \& {Lee}(1984)}]{1984ApJ...285...89D}
{Draine}, B.~T., \& {Lee}, H.~M. 1984, \apj, 285, 89, \dodoi{10.1086/162480}

\bibitem[{{Draine} \& {Li}(2007)}]{2007ApJ...657..810D}
{Draine}, B.~T., \& {Li}, A. 2007, \apj, 657, 810, \dodoi{10.1086/511055}

\bibitem[{{Draine} {et~al.}(2014){Draine}, {Aniano}, {Krause}, {Groves},
  {Sandstrom}, {Braun}, {Leroy}, {Klaas}, {Linz}, {Rix}, {Schinnerer},
  {Schmiedeke}, \& {Walter}}]{2014ApJ...780..172D}
{Draine}, B.~T., {Aniano}, G., {Krause}, O., {et~al.} 2014, \apj, 780, 172,
  \dodoi{10.1088/0004-637X/780/2/172}

\bibitem[{{Fitzpatrick}(1986)}]{1986AJ.....92.1068F}
{Fitzpatrick}, E.~L. 1986, \aj, 92, 1068, \dodoi{10.1086/114237}

\bibitem[{{Fitzpatrick}(1999)}]{1999PASP..111...63F}
---. 1999, \pasp, 111, 63, \dodoi{10.1086/316293}

\bibitem[{{Fitzpatrick} \& {Massa}(1990)}]{1990ApJS...72..163F}
{Fitzpatrick}, E.~L., \& {Massa}, D. 1990, \apjs, 72, 163,
  \dodoi{10.1086/191413}

\bibitem[{{Fitzpatrick} \& {Massa}(2005)}]{2005AJ....130.1127F}
---. 2005, \aj, 130, 1127, \dodoi{10.1086/431900}

\bibitem[{{Foreman-Mackey} {et~al.}(2013){Foreman-Mackey}, {Hogg}, {Lang}, \&
  {Goodman}}]{2013PASP..125..306F}
{Foreman-Mackey}, D., {Hogg}, D.~W., {Lang}, D., \& {Goodman}, J. 2013, \pasp,
  125, 306, \dodoi{10.1086/670067}

\bibitem[{{Galliano} {et~al.}(2018){Galliano}, {Galametz}, \&
  {Jones}}]{2018ARA&A..56..673G}
{Galliano}, F., {Galametz}, M., \& {Jones}, A.~P. 2018, \araa, 56, 673,
  \dodoi{10.1146/annurev-astro-081817-051900}

\bibitem[{{Gao} {et~al.}(2015){Gao}, {Jiang}, {Li}, {Li}, \&
  {Wang}}]{2015ApJ...807L..26G}
{Gao}, J., {Jiang}, B.~W., {Li}, A., {Li}, J., \& {Wang}, X. 2015, \apjl, 807,
  L26, \dodoi{10.1088/2041-8205/807/2/L26}

\bibitem[{{Gao} {et~al.}(2013){Gao}, {Li}, \& {Jiang}}]{2013EP&S...65.1127G}
{Gao}, J., {Li}, A., \& {Jiang}, B.~W. 2013, Earth, Planets, and Space, 65,
  1127, \dodoi{10.5047/eps.2013.05.016}

\bibitem[{{Gao} {et~al.}(2020){Gao}, {Zhao}, {Gao}, {Jiang}, \&
  {Li}}]{2020P&SS..18304627G}
{Gao}, W., {Zhao}, R., {Gao}, J., {Jiang}, B., \& {Li}, J. 2020, \planss, 183,
  104627, \dodoi{10.1016/j.pss.2018.12.010}

\bibitem[{{Gordon} {et~al.}(2003){Gordon}, {Clayton}, {Misselt}, {Land olt}, \&
  {Wolff}}]{2003ApJ...594..279G}
{Gordon}, K.~D., {Clayton}, G.~C., {Misselt}, K.~A., {Land olt}, A.~U., \&
  {Wolff}, M.~J. 2003, \apj, 594, 279, \dodoi{10.1086/376774}

\bibitem[{{Gordon} {et~al.}(2016){Gordon}, {Fouesneau}, {Arab}, {Tchernyshyov},
  {Weisz}, {Dalcanton}, {Williams}, {Bell}, {Bianchi}, {Boyer}, {Choi},
  {Dolphin}, {Girardi}, {Hogg}, {Kalirai}, {Kapala}, {Lewis}, {Rix},
  {Sandstrom}, \& {Skillman}}]{2016ApJ...826..104G}
{Gordon}, K.~D., {Fouesneau}, M., {Arab}, H., {et~al.} 2016, \apj, 826, 104,
  \dodoi{10.3847/0004-637X/826/2/104}

\bibitem[{{Green} {et~al.}(2019){Green}, {Schlafly}, {Zucker}, {Speagle}, \&
  {Finkbeiner}}]{2019ApJ...887...93G}
{Green}, G.~M., {Schlafly}, E., {Zucker}, C., {Speagle}, J.~S., \&
  {Finkbeiner}, D. 2019, \apj, 887, 93, \dodoi{10.3847/1538-4357/ab5362}

\bibitem[{{Guo} {et~al.}(2021){Guo}, {Chen}, {Yuan}, {Huang}, {Liu}, {Yang},
  {Li}, {Sun}, \& {Liu}}]{2021ApJ...906...47G}
{Guo}, H.~L., {Chen}, B.~Q., {Yuan}, H.~B., {et~al.} 2021, \apj, 906, 47,
  \dodoi{10.3847/1538-4357/abc68a}

\bibitem[{{Hodapp} {et~al.}(2004){Hodapp}, {Kaiser}, {Aussel}, {Burgett},
  {Chambers}, {Chun}, {Dombeck}, {Douglas}, {Hafner}, {Heasley}, {Hoblitt},
  {Hude}, {Isani}, {Jedicke}, {Jewitt}, {Laux}, {Luppino}, {Lupton}, {Maberry},
  {Magnier}, {Mannery}, {Monet}, {Morgan}, {Onaka}, {Price}, {Ryan},
  {Siegmund}, {Szapudi}, {Tonry}, {Wainscoat}, \&
  {Waterson}}]{2004AN....325..636H}
{Hodapp}, K.~W., {Kaiser}, N., {Aussel}, H., {et~al.} 2004, Astronomische
  Nachrichten, 325, 636, \dodoi{10.1002/asna.200410300}

\bibitem[{{Hubeny}(1988)}]{1988CoPhC..52..103H}
{Hubeny}, I. 1988, Computer Physics Communications, 52, 103,
  \dodoi{10.1016/0010-4655(88)90177-4}

\bibitem[{{Hubeny} \& {Lanz}(2017{\natexlab{a}})}]{2017arXiv170601859H}
{Hubeny}, I., \& {Lanz}, T. 2017{\natexlab{a}}, arXiv e-prints,
  arXiv:1706.01859.
\newblock \doarXiv{1706.01859}

\bibitem[{{Hubeny} \& {Lanz}(2017{\natexlab{b}})}]{2017arXiv170601937H}
---. 2017{\natexlab{b}}, arXiv e-prints, arXiv:1706.01937.
\newblock \doarXiv{1706.01937}

\bibitem[{{Hubeny} \& {Lanz}(2017{\natexlab{c}})}]{2017arXiv170601935H}
---. 2017{\natexlab{c}}, arXiv e-prints, arXiv:1706.01935.
\newblock \doarXiv{1706.01935}

\bibitem[{{Huffman} \& {Stapp}(1973)}]{1973IAUS...52..297H}
{Huffman}, D.~R., \& {Stapp}, J.~L. 1973, in Interstellar Dust and Related
  Topics, ed. J.~M. {Greenberg} \& H.~C. {van de Hulst}, Vol.~52, 297

\bibitem[{{Irwin}(2013)}]{2013ASSP...37..229I}
{Irwin}, M.~J. 2013, in Thirty Years of Astronomical Discovery with UKIRT,
  Vol.~37, 229, \dodoi{10.1007/978-94-007-7432-2_21}

\bibitem[{{Joblin} {et~al.}(1992){Joblin}, {Leger}, \&
  {Martin}}]{1992ApJ...393L..79J}
{Joblin}, C., {Leger}, A., \& {Martin}, P. 1992, \apjl, 393, L79,
  \dodoi{10.1086/186456}

\bibitem[{{Joseph} {et~al.}(1984){Joseph}, {Meikle}, {Robertson}, \&
  {Wright}}]{1984MNRAS.209..111J}
{Joseph}, R.~D., {Meikle}, W.~P.~S., {Robertson}, N.~A., \& {Wright}, G.~S.
  1984, \mnras, 209, 111, \dodoi{10.1093/mnras/209.1.111}

\bibitem[{{Kim} {et~al.}(1994){Kim}, {Martin}, \&
  {Hendry}}]{1994ApJ...422..164K}
{Kim}, S.-H., {Martin}, P.~G., \& {Hendry}, P.~D. 1994, \apj, 422, 164,
  \dodoi{10.1086/173714}

\bibitem[{{Kurucz}(1970)}]{1970SAOSR.309.....K}
{Kurucz}, R.~L. 1970, SAO Special Report, 309

\bibitem[{{Lanz} \& {Hubeny}(2003)}]{2003ApJS..146..417L}
{Lanz}, T., \& {Hubeny}, I. 2003, \apjs, 146, 417, \dodoi{10.1086/374373}

\bibitem[{{Lanz} \& {Hubeny}(2007)}]{2007ApJS..169...83L}
---. 2007, \apjs, 169, 83, \dodoi{10.1086/511270}

\bibitem[{{Lequeux} {et~al.}(1982){Lequeux}, {Maurice}, {Prevot-Burnichon},
  {Prevot}, \& {Rocca-Volmerange}}]{1982A&A...113L..15L}
{Lequeux}, J., {Maurice}, E., {Prevot-Burnichon}, M.~L., {Prevot}, L., \&
  {Rocca-Volmerange}, B. 1982, \aap, 113, L15

\bibitem[{{Li}(2005)}]{2005ApJ...622..965L}
{Li}, A. 2005, \apj, 622, 965, \dodoi{10.1086/428038}

\bibitem[{{Li} \& {Draine}(2001)}]{2001ApJ...554..778L}
{Li}, A., \& {Draine}, B.~T. 2001, \apj, 554, 778, \dodoi{10.1086/323147}

\bibitem[{{Liu} {et~al.}(2019){Liu}, {Cui}, {Liu}, {Huang}, {Zhao}, \&
  {Zhang}}]{2019ApJS..241...32L}
{Liu}, Z., {Cui}, W., {Liu}, C., {et~al.} 2019, \apjs, 241, 32,
  \dodoi{10.3847/1538-4365/ab0a0d}

\bibitem[{{Martins} \& {Plez}(2006)}]{2006A&A...457..637M}
{Martins}, F., \& {Plez}, B. 2006, \aap, 457, 637,
  \dodoi{10.1051/0004-6361:20065753}

\bibitem[{{Maryeva} {et~al.}(2016{\natexlab{a}}){Maryeva}, {Chentsov},
  {Goranskij}, {Dyachenko}, {Karpov}, {Malogolovets}, \&
  {Rastegaev}}]{2016MNRAS.458..491M}
{Maryeva}, O.~V., {Chentsov}, E.~L., {Goranskij}, V.~P., {et~al.}
  2016{\natexlab{a}}, \mnras, 458, 491, \dodoi{10.1093/mnras/stw385}

\bibitem[{{Maryeva} {et~al.}(2016{\natexlab{b}}){Maryeva}, {Chentsov},
  {Goranskij}, \& {Karpov}}]{2016BaltA..25...42M}
{Maryeva}, O.~V., {Chentsov}, E.~L., {Goranskij}, V.~P., \& {Karpov}, S.~V.
  2016{\natexlab{b}}, Baltic Astronomy, 25, 42, \dodoi{10.1515/astro-2017-0108}

\bibitem[{{Mason} {et~al.}(2001){Mason}, {Breeveld}, {Much}, {Carter},
  {Cordova}, {Cropper}, {Fordham}, {Huckle}, {Ho}, {Kawakami}, {Kennea},
  {Kennedy}, {Mittaz}, {Pandel}, {Priedhorsky}, {Sasseen}, {Shirey}, {Smith},
  \& {Vreux}}]{2001A&A...365L..36M}
{Mason}, K.~O., {Breeveld}, A., {Much}, R., {et~al.} 2001, \aap, 365, L36,
  \dodoi{10.1051/0004-6361:20000044}

\bibitem[{{Massey} {et~al.}(2007{\natexlab{a}}){Massey}, {McNeill}, {Olsen},
  {Hodge}, {Blaha}, {Jacoby}, {Smith}, \& {Strong}}]{2007AJ....134.2474M}
{Massey}, P., {McNeill}, R.~T., {Olsen}, K.~A.~G., {et~al.} 2007{\natexlab{a}},
  \aj, 134, 2474, \dodoi{10.1086/523658}

\bibitem[{{Massey} {et~al.}(2016){Massey}, {Neugent}, \&
  {Smart}}]{2016AJ....152...62M}
{Massey}, P., {Neugent}, K.~F., \& {Smart}, B.~M. 2016, \aj, 152, 62,
  \dodoi{10.3847/0004-6256/152/3/62}

\bibitem[{{Massey} {et~al.}(2007{\natexlab{b}}){Massey}, {Olsen}, {Hodge},
  {Jacoby}, {McNeill}, {Smith}, \& {Strong}}]{2007AJ....133.2393M}
{Massey}, P., {Olsen}, K.~A.~G., {Hodge}, P.~W., {et~al.} 2007{\natexlab{b}},
  \aj, 133, 2393, \dodoi{10.1086/513319}

\bibitem[{{Massey} {et~al.}(2011){Massey}, {Olsen}, {Hodge}, {Jacoby},
  {McNeill}, {Smith}, \& {Strong}}]{2011AJ....141...28M}
---. 2011, \aj, 141, 28, \dodoi{10.1088/0004-6256/141/1/28}

\bibitem[{{Massey} {et~al.}(2006){Massey}, {Olsen}, {Hodge}, {Strong},
  {Jacoby}, {Schlingman}, \& {Smith}}]{2006AJ....131.2478M}
---. 2006, \aj, 131, 2478, \dodoi{10.1086/503256}

\bibitem[{{Mathis}(1990)}]{1990ARA&A..28...37M}
{Mathis}, J.~S. 1990, \araa, 28, 37,
  \dodoi{10.1146/annurev.aa.28.090190.000345}

\bibitem[{{Mathis} {et~al.}(1977){Mathis}, {Rumpl}, \&
  {Nordsieck}}]{1977ApJ...217..425M}
{Mathis}, J.~S., {Rumpl}, W., \& {Nordsieck}, K.~H. 1977, \apj, 217, 425,
  \dodoi{10.1086/155591}

\bibitem[{{McConnachie} {et~al.}(2005){McConnachie}, {Irwin}, {Ferguson},
  {Ibata}, {Lewis}, \& {Tanvir}}]{2005MNRAS.356..979M}
{McConnachie}, A.~W., {Irwin}, M.~J., {Ferguson}, A.~M.~N., {et~al.} 2005,
  \mnras, 356, 979, \dodoi{10.1111/j.1365-2966.2004.08514.x}

\bibitem[{{Mishra} \& {Li}(2015)}]{2015ApJ...809..120M}
{Mishra}, A., \& {Li}, A. 2015, \apj, 809, 120,
  \dodoi{10.1088/0004-637X/809/2/120}

\bibitem[{{Mishra} \& {Li}(2017)}]{2017ApJ...850..138M}
---. 2017, \apj, 850, 138, \dodoi{10.3847/1538-4357/aa937a}

\bibitem[{{Nandy} \& {Morgan}(1978)}]{1978Natur.276..478N}
{Nandy}, K., \& {Morgan}, D.~H. 1978, \nat, 276, 478, \dodoi{10.1038/276478a0}

\bibitem[{{Neugent} {et~al.}(2020){Neugent}, {Massey}, {Georgy}, {Drout},
  {Mommert}, {Levesque}, {Meynet}, \& {Ekstr{\"o}m}}]{2020ApJ...889...44N}
{Neugent}, K.~F., {Massey}, P., {Georgy}, C., {et~al.} 2020, \apj, 889, 44,
  \dodoi{10.3847/1538-4357/ab5ba0}

\bibitem[{{Page} {et~al.}(2019){Page}, {Brindle}, {Talavera}, {Still}, {Rosen},
  {Yershov}, {Ziaeepour}, {Mason}, {Cropper}, {Breeveld}, {Loiseau}, {Mignani},
  {Smith}, \& {Murdin}}]{2019yCat.2356....0P}
{Page}, M.~J., {Brindle}, C., {Talavera}, A., {et~al.} 2019, VizieR Online Data
  Catalog, II/356

\bibitem[{{Peek} \& {Schiminovich}(2013)}]{2013ApJ...771...68P}
{Peek}, J.~E.~G., \& {Schiminovich}, D. 2013, \apj, 771, 68,
  \dodoi{10.1088/0004-637X/771/1/68}

\bibitem[{{Prevot} {et~al.}(1984){Prevot}, {Lequeux}, {Maurice}, {Prevot}, \&
  {Rocca-Volmerange}}]{1984A&A...132..389P}
{Prevot}, M.~L., {Lequeux}, J., {Maurice}, E., {Prevot}, L., \&
  {Rocca-Volmerange}, B. 1984, \aap, 132, 389

\bibitem[{{Ren} {et~al.}(2021){Ren}, {Jiang}, {Yang}, {Wang}, {Jian}, \&
  {Ren}}]{2021ApJ...907...18R}
{Ren}, Y., {Jiang}, B., {Yang}, M., {et~al.} 2021, \apj, 907, 18,
  \dodoi{10.3847/1538-4357/abcda5}

\bibitem[{{Rodrigo} {et~al.}(2012){Rodrigo}, {Solano}, \&
  {Bayo}}]{2012ivoa.rept.1015R}
{Rodrigo}, C., {Solano}, E., \& {Bayo}, A. 2012, {SVO Filter Profile Service
  Version 1.0}, IVOA Working Draft 15 October 2012,
  \dodoi{10.5479/ADS/bib/2012ivoa.rept.1015R}

\bibitem[{{Roming} {et~al.}(2005){Roming}, {Kennedy}, {Mason}, {Nousek}, {Ahr},
  {Bingham}, {Broos}, {Carter}, {Hancock}, {Huckle}, {Hunsberger}, {Kawakami},
  {Killough}, {Koch}, {McLelland}, {Smith}, {Smith}, {Soto}, {Boyd},
  {Breeveld}, {Holland}, {Ivanushkina}, {Pryzby}, {Still}, \&
  {Stock}}]{2005SSRv..120...95R}
{Roming}, P. W.~A., {Kennedy}, T.~E., {Mason}, K.~O., {et~al.} 2005, \ssr, 120,
  95, \dodoi{10.1007/s11214-005-5095-4}

\bibitem[{{Ruoyi} \& {Haibo}(2020)}]{2020ApJ...905L..20R}
{Ruoyi}, Z., \& {Haibo}, Y. 2020, \apjl, 905, L20,
  \dodoi{10.3847/2041-8213/abccc4}

\bibitem[{{Sale} {et~al.}(2014){Sale}, {Drew}, {Barentsen}, {Farnhill},
  {Raddi}, {Barlow}, {Eisl{\"o}ffel}, {Vink}, {Rodr{\'\i}guez-Gil}, \&
  {Wright}}]{2014MNRAS.443.2907S}
{Sale}, S.~E., {Drew}, J.~E., {Barentsen}, G., {et~al.} 2014, \mnras, 443,
  2907, \dodoi{10.1093/mnras/stu1090}

\bibitem[{{Sanders} {et~al.}(2012){Sanders}, {Caldwell}, {McDowell}, \&
  {Harding}}]{2012ApJ...758..133S}
{Sanders}, N.~E., {Caldwell}, N., {McDowell}, J., \& {Harding}, P. 2012, \apj,
  758, 133, \dodoi{10.1088/0004-637X/758/2/133}

\bibitem[{{Schlafly} \& {Finkbeiner}(2011)}]{2011ApJ...737..103S}
{Schlafly}, E.~F., \& {Finkbeiner}, D.~P. 2011, \apj, 737, 103,
  \dodoi{10.1088/0004-637X/737/2/103}

\bibitem[{{Schlafly} {et~al.}(2012){Schlafly}, {Finkbeiner}, {Juri{\'c}},
  {Magnier}, {Burgett}, {Chambers}, {Grav}, {Hodapp}, {Kaiser}, {Kudritzki},
  {Martin}, {Morgan}, {Price}, {Rix}, {Stubbs}, {Tonry}, \&
  {Wainscoat}}]{2012ApJ...756..158S}
{Schlafly}, E.~F., {Finkbeiner}, D.~P., {Juri{\'c}}, M., {et~al.} 2012, \apj,
  756, 158, \dodoi{10.1088/0004-637X/756/2/158}

\bibitem[{{Schlegel} {et~al.}(1998){Schlegel}, {Finkbeiner}, \&
  {Davis}}]{1998ApJ...500..525S}
{Schlegel}, D.~J., {Finkbeiner}, D.~P., \& {Davis}, M. 1998, \apj, 500, 525,
  \dodoi{10.1086/305772}

\bibitem[{{Shao} {et~al.}(2018){Shao}, {Jiang}, {Li}, {Gao}, {Lv}, \&
  {Yao}}]{2018MNRAS.478.3467S}
{Shao}, Z., {Jiang}, B.~W., {Li}, A., {et~al.} 2018, \mnras, 478, 3467,
  \dodoi{10.1093/mnras/sty1267}

\bibitem[{{Smartt} {et~al.}(2001){Smartt}, {Crowther}, {Dufton}, {Lennon},
  {Kudritzki}, {Herrero}, {McCarthy}, \& {Bresolin}}]{2001MNRAS.325..257S}
{Smartt}, S.~J., {Crowther}, P.~A., {Dufton}, P.~L., {et~al.} 2001, \mnras,
  325, 257, \dodoi{10.1046/j.1365-8711.2001.04415.x}

\bibitem[{{Smith} \& {Hancock}(2009)}]{2009AJ....138..130S}
{Smith}, B.~J., \& {Hancock}, M. 2009, \aj, 138, 130,
  \dodoi{10.1088/0004-6256/138/1/130}

\bibitem[{{Steglich} {et~al.}(2011){Steglich}, {Bouwman}, {Huisken}, \&
  {Henning}}]{2011ApJ...742....2S}
{Steglich}, M., {Bouwman}, J., {Huisken}, F., \& {Henning}, T. 2011, \apj, 742,
  2, \dodoi{10.1088/0004-637X/742/1/2}

\bibitem[{{Stubbs} {et~al.}(2010){Stubbs}, {Doherty}, {Cramer}, {Narayan},
  {Brown}, {Lykke}, {Woodward}, \& {Tonry}}]{2010ApJS..191..376S}
{Stubbs}, C.~W., {Doherty}, P., {Cramer}, C., {et~al.} 2010, \apjs, 191, 376,
  \dodoi{10.1088/0067-0049/191/2/376}

\bibitem[{{Tonry} {et~al.}(2012){Tonry}, {Stubbs}, {Lykke}, {Doherty},
  {Shivvers}, {Burgett}, {Chambers}, {Hodapp}, {Kaiser}, {Kudritzki},
  {Magnier}, {Morgan}, {Price}, \& {Wainscoat}}]{2012ApJ...750...99T}
{Tonry}, J.~L., {Stubbs}, C.~W., {Lykke}, K.~R., {et~al.} 2012, \apj, 750, 99,
  \dodoi{10.1088/0004-637X/750/2/99}

\bibitem[{{Van De Putte} {et~al.}(2020){Van De Putte}, {Gordon}, {Roman-Duval},
  {Williams}, {Baes}, {Tchernyshyov}, {Lawton}, \&
  {Arab}}]{2020ApJ...888...22V}
{Van De Putte}, D., {Gordon}, K.~D., {Roman-Duval}, J., {et~al.} 2020, \apj,
  888, 22, \dodoi{10.3847/1538-4357/ab557f}

\bibitem[{{Vanzi} {et~al.}(2000){Vanzi}, {Hunt}, {Thuan}, \&
  {Izotov}}]{2000A&A...363..493V}
{Vanzi}, L., {Hunt}, L.~K., {Thuan}, T.~X., \& {Izotov}, Y.~I. 2000, \aap, 363,
  493.
\newblock \doarXiv{astro-ph/0009218}

\bibitem[{{Venn} {et~al.}(2000){Venn}, {McCarthy}, {Lennon}, {Przybilla},
  {Kudritzki}, \& {Lemke}}]{2000ApJ...541..610V}
{Venn}, K.~A., {McCarthy}, J.~K., {Lennon}, D.~J., {et~al.} 2000, \apj, 541,
  610, \dodoi{10.1086/309491}

\bibitem[{{Wang} \& {Chen}(2019)}]{2019ApJ...877..116W}
{Wang}, S., \& {Chen}, X. 2019, \apj, 877, 116,
  \dodoi{10.3847/1538-4357/ab1c61}

\bibitem[{{Wang} {et~al.}(2017){Wang}, {Jiang}, {Zhao}, {Chen}, \& {de
  Grijs}}]{2017ApJ...848..106W}
{Wang}, S., {Jiang}, B.~W., {Zhao}, H., {Chen}, X., \& {de Grijs}, R. 2017,
  \apj, 848, 106, \dodoi{10.3847/1538-4357/aa8db7}

\bibitem[{{Wang} {et~al.}(2014){Wang}, {Li}, \& {Jiang}}]{2014P&SS..100...32W}
{Wang}, S., {Li}, A., \& {Jiang}, B.~W. 2014, \planss, 100, 32,
  \dodoi{10.1016/j.pss.2014.03.018}

\bibitem[{{Wegner}(1994)}]{1994MNRAS.270..229W}
{Wegner}, W. 1994, \mnras, 270, 229, \dodoi{10.1093/mnras/270.2.229}

\bibitem[{{Weingartner} \& {Draine}(2001)}]{2001ApJ...548..296W}
{Weingartner}, J.~C., \& {Draine}, B.~T. 2001, \apj, 548, 296,
  \dodoi{10.1086/318651}

\bibitem[{{Welty} \& {Fowler}(1992)}]{1992ApJ...393..193W}
{Welty}, D.~E., \& {Fowler}, J.~R. 1992, \apj, 393, 193, \dodoi{10.1086/171497}

\bibitem[{{Williams} {et~al.}(2014){Williams}, {Lang}, {Dalcanton}, {Dolphin},
  {Weisz}, {Bell}, {Bianchi}, {Byler}, {Gilbert}, {Girardi}, {Gordon},
  {Gregersen}, {Johnson}, {Kalirai}, {Lauer}, {Monachesi}, {Rosenfield},
  {Seth}, \& {Skillman}}]{2014ApJS..215....9W}
{Williams}, B.~F., {Lang}, D., {Dalcanton}, J.~J., {et~al.} 2014, \apjs, 215,
  9, \dodoi{10.1088/0067-0049/215/1/9}

\bibitem[{{Wright} {et~al.}(2015){Wright}, {Drew}, \&
  {Mohr-Smith}}]{2015MNRAS.449..741W}
{Wright}, N.~J., {Drew}, J.~E., \& {Mohr-Smith}, M. 2015, \mnras, 449, 741,
  \dodoi{10.1093/mnras/stv323}

\bibitem[{{Xiang} {et~al.}(2011){Xiang}, {Li}, \&
  {Zhong}}]{2011ApJ...733...91X}
{Xiang}, F.~Y., {Li}, A., \& {Zhong}, J.~X. 2011, \apj, 733, 91,
  \dodoi{10.1088/0004-637X/733/2/91}

\bibitem[{{Yershov}(2015)}]{2015yCat.2339....0Y}
{Yershov}, V.~N. 2015, VizieR Online Data Catalog, II/339

\bibitem[{{Zaritsky} {et~al.}(1994){Zaritsky}, {Kennicutt}, \&
  {Huchra}}]{1994ApJ...420...87Z}
{Zaritsky}, D., {Kennicutt}, Robert~C., J., \& {Huchra}, J.~P. 1994, \apj, 420,
  87, \dodoi{10.1086/173544}

\bibitem[{{Zhan}(2021)}]{2021CSB...111.111C}
{Zhan}, H. 2021, Chinese Science Bulletin, ,
  \dodoi{https://doi.org/10.1360/TB-2021-0016}

\end{thebibliography}
\bibliographystyle{aasjournal}



\end{CJK*}
\end{document}